\documentclass[twocolumn,aps,prd,preprintnumbers,groupedaddress,tightenlines,nofootinbib,floatfix,10pt,reprint,superscriptaddress]{revtex4-2}

\usepackage{amssymb,amsmath,bm,graphicx,hyperref,xcolor}

\hypersetup{
    colorlinks=true,       
    linkcolor=blue,          
    citecolor=blue,        
    urlcolor=blue,           
}

\begin{document}

\preprint{YITP-24-87, J-PARC-TH-0307, UTHEP-790}

\title{High-precision analysis of the critical point in heavy-quark QCD at $N_t=6$
}

\author{Ryo Ashikawa}
\affiliation{
  Department of Physics, Osaka University, Toyonaka, Osaka 560-0043, Japan}
\author{Masakiyo Kitazawa}
\email[]{kitazawa@yukawa.kyoto-u.ac.jp}
\affiliation{
  Yukawa Institute for Theoretical Physics, Kyoto University, Kyoto, 606-8502, Japan}
\affiliation{
  J-PARC Branch, KEK Theory Center, 
  Institute of Particle and Nuclear Studies, KEK, Tokai, Ibaraki 319-1106, Japan}
\author{Shinji Ejiri}
\affiliation{
  Department of Physics, Niigata University, Niigata 950-2181, Japan}
\author{Kazuyuki Kanaya}
\affiliation{
  Tomonaga Center for the History of the Universe, University of Tsukuba, Tsukuba, Ibaraki 305-8571, Japan}

\date{\today}

\begin{abstract}

Binder-cumulant analysis of the critical point in the heavy-quark region of QCD is performed by Monte-Carlo simulations with the hopping-parameter expansion at $N_t=6$. We extend our previous analysis at $N_t=4$ to finer lattices and perform high-precision analyses on large spatial volumes up to the aspect ratio $LT=N_s/N_t=18$. Higher order terms in the hopping-parameter expansion are incorporated effectively up to 14th order. The numerical results show that the violation of the finite-size scaling becomes more prominent on the finer lattice at a given aspect ratio.

\end{abstract}

\maketitle

\section{Introduction}
\label{sec:intro}

Appearance of critical points (CPs) in Quantum Chromodynamics (QCD) in medium with variations of various external parameters is one of the interesting features of this theory. CPs are expected to exist on the phase diagram of QCD on the temperature and baryon chemical potential plane at physical quark masses~\cite{Asakawa:1989bq,Kitazawa:2002jop}. These CPs are the main targets of the beam-energy scan program in relativistic heavy-ion collisions~\cite{Asakawa:2015ybt,Bluhm:2020mpc,STAR:2021iop,STAR:2022etb}. Although their search in lattice-QCD numerical simulations is difficult due to the sign problem, various attempts to approach the CP are ongoing~\cite{HotQCD:2018pds,Borsanyi:2020fev,Scherzer:2020kiu,Guenther:2020jwe,Fukuma:2020fez,Dini:2021hug,Dimopoulos:2021vrk,Borsanyi:2021hbk,Namekawa:2022liz,Aarts:2023vsf,Yokota:2023osv,Clarke:2024ugt}. 
CPs are also expected to manifest themselves at vanishing baryon chemical potential when the quark masses are varied from the physical ones both toward the light and heavy directions, as nicely summarized on the so-called Columbia plot~\cite{Pisarski:1983ms,Philipsen:2021qji}. 
Even other CPs may manifest themselves provoked by spatial boundary conditions~\cite{Fujii:2024llh}.

The CPs in the light and heavy quark-mass regions have been investigated in lattice-QCD numerical simulations for decades~\cite{Saito:2013vja,Jin:2013wta,Ejiri:2015vip,Bazavov:2017xul,Jin:2017jjp,Ejiri:2019csa,Philipsen:2019rjq,Kuramashi:2020meg,Cuteri:2020yke,Kiyohara:2021smr,Dini:2021hug,Zhang:2022kzb}. Since they appear at zero baryon chemical potential, these analyses can be carried out without suffering from the sign problem. 
For the CP in the light-quark region, however, recent analyses give controversial results and even the existence of the CP off the chiral limit is not clear~\cite{Kuramashi:2020meg,Philipsen:2021qji,Dini:2021hug}. While the existence of the CP is robust for the heavy-quark region, a proper reproduction of the expected scaling behavior from the $Z(2)$ universality class, to which the CP is believed to belong, is known to be difficult~\cite{Cuteri:2020yke,Kiyohara:2021smr}. 

The Binder-cumulant analysis~\cite{Binder:1981sa} is one of the useful methods that have been used in these studies. Using the finite-size scaling (FSS) of the scaling function, this method allows us to reveal various properties of the CP in the thermodynamic limit, such as the location and critical exponents, from the numerical results for finite-volume systems. 
In the Binder-cumulant analysis, however, it is usually assumed that the thermodynamics is dominated by the ``singular part'' that obeys the scaling behavior. This assumption, however, is justified only in the vicinity of the CP and for sufficiently large spatial volumes.
In practical numerical simulations that are performed on finite volumes, the numerical results are always contaminated by the ``non-singular part'' that violates the ideal FSS behavior. The controversial situations in the analyses both in the light and heavy quark-mass regions may stem from the non-singular part that is not suppressed well at the spatial volumes used in the simulations.

In Ref.~\cite{Kiyohara:2021smr}, motivated by this observation, part of the authors of the present paper have performed the Binder-cumulant analysis around the CP in the heavy-quark region of QCD on large spatial volumes. The aspect ratio is taken up to $LT=N_s/N_t=12$, with $L=N_s a$ the spatial lattice size and $T=1/(N_t a)$ the temperature, while the lattice spacing $a$ was fixed to a coarse one with the temporal lattice extent $N_t=4$. To realize the large-volume simulations, the hopping-parameter expansion (HPE) for the quark determinant was employed: Monte-Carlo simulations are performed at the leading non-trivial order (LO) of the HPE, and the next-to-leading-order (NLO) contributions are taken into account by reweighting. 
It has been shown that the truncation error of the HPE is well suppressed around the CP at $N_t=4$ in this analysis, and also this method enables efficient numerical analysis there. 
Through the Binder-cumulant analysis using the precision numerical results on large lattices, it was found that the violation of the FSS is not negligible even at $LT=8$ for $N_t=4$. 

In the present study, we extend this analysis to a finer lattice with $N_t=6$. 
To see the correct scaling behavior, we also enlarge the aspect ratio up to $LT=18$.
Besides the prescription to deal with the HPE up to NLO used in Ref.~\cite{Kiyohara:2021smr}, we newly employ a method proposed in Ref.~\cite{Wakabayashi:2021eye} to incorporate higher order terms in the HPE effectively into the analysis. 
Although the convergence of the HPE becomes worse as the continuum limit is approached, this method allows us to incorporate virtually all orders in the HPE near the CP at $N_t=6$.
These numerical results enable us to fix the location of the CP with high precision. It will also be shown that the violation of the FSS becomes more prominent at $N_t=6$ compared to the coarse-lattice simulations in Ref.~\cite{Kiyohara:2021smr}.

This paper is organized as follows. In the next section, we describe the lattice action and its HPE, and the procedure to deal with the higher-order terms of the HPE in the numerical analysis. In Sec.~\ref{sec:detail}, we discuss the detailed setup and parameters of our Monte-Carlo simulations. The numerical results are then presented in Sec.~\ref{sec:result}. The last section is devoted to a short summary. The convergence of the HPE is discussed in Appendix~\ref{sec:convergence}.

\section{Hopping-parameter expansion}
\label{sec:setup}

In this section, we summarize the HPE and a method to handle the effects of its higher-order terms.

\subsection{Lattice action}
\label{sec:action}

Our lattice QCD action is decomposed as $S = S_{\rm g} + S_{\rm q}$ with the gauge and quark actions $S_{\rm g}$ and $S_{\rm q}$. We employ the Wilson gauge action
\begin{align}
S_{\rm g} = -6 N_{\rm site} \,\beta \, \hat{P},
\label{eq:Sg}
\end{align} 
with the space-time lattice volume $N_{\rm site} = N_s^3 \times N_t$, the gauge coupling parameter $\beta = 6/g^2$, and the plaquette operator
\begin{align}
\hat{P}= \frac{1}{6 N_{\rm c} N_{\rm site}} \displaystyle \sum_{x,\,\mu < \nu} 
 {\rm Re \ tr_C} \left[ U_{x,\mu} U_{x+\hat{\mu},\nu}
U^{\dagger}_{x+\hat{\nu},\mu} U^{\dagger}_{x,\nu} \right] ,
\label{eq:Plaq}
\end{align} 
where $U_{x,\mu}$ is the link variable in the $\mu$ direction at site $x$, 
${\rm tr_C}$ is the trace over color indices, and $N_{\rm c}=3$.
For $S_{\rm q}$, we employ the Wilson quark action 
\begin{align}
  S_{\rm q} = \sum_{f=1}^{N_{\rm f}} \sum_{x,\,y} \bar{\psi}_x^{(f)} \,
  M_{xy} (\kappa_f) \, \psi_y^{(f)} ,
\label{eq:Sq}
\end{align} 
with the Wilson quark kernel
\begin{align}
  &M_{xy} (\kappa) = \delta_{xy} - \kappa B_{xy},
  \label{eq:Mxy}
  \\
  &B_{xy}
  =  \sum_{\mu=1}^4 \left[ (1-\gamma_{\mu})\,U_{x,\mu}\,\delta_{y,x+\hat{\mu}} + (1+\gamma_{\mu})\,U_{y,\mu}^{\dagger}\,\delta_{y,x-\hat{\mu}} \right],
  \label{eq:B}
\end{align} 
where the color and Dirac-spinor indices are suppressed for simplicity.
$\kappa_f$ is the hopping parameter for the $f$th flavor, which is related to the bare quark mass $m_f$ as $\kappa_f=1/(2m_fa+8)$. 
In the following, we consider the case of degenerated $N_{\rm f}$ flavors with a common hopping parameter $\kappa=\kappa_f$, whereas generalization to non-degenerate cases is straightforward.

After integrating out the quark degrees of freedom, the expectation value of a gauge operator $\hat{O}(U)$ is given by
\begin{align}
  \langle \hat{O}(U) \rangle
  &= \frac1Z \int {\cal D}U \, \hat{O}(U) \, e^{-S_{\rm eff}},  
  \label{eq:<O>}
\end{align}
with 
\begin{align}
  S_{\rm eff} = S_{\rm g} - N_{\rm f} \ln{\rm det}M(\kappa) ,
  \label{eq:S_eff}
\end{align}
and the partition function $Z=\int {\cal D}U \, e^{-S_{\rm eff}}$.

\subsection{Hopping-parameter expansion}
\label{sec:HPE}

In the heavy-quark region $\kappa\ll1$, $\ln{\rm det}M$ in Eq.~(\ref{eq:S_eff}) is expanded with respect to $\kappa$ at $\kappa=0$ as 
\begin{align}
  \ln \left[ \frac{\det M(\kappa)}{\det M(0)} \right]
  =-\sum_{n=1}^{\infty} \frac1n {\rm Tr} 
  \left[ B^n \right] \kappa^n .
  \label{eq:HPE}
\end{align}
Here, ${\rm Tr}$ is the trace over all indices
and an irrelevant contribution at $\kappa=0$ is subtracted in Eq.~(\ref{eq:HPE}).
The $n$th-order terms in the HPE are graphically represented by the closed trajectories of length $n$~\cite{Rothe:1992nt,Kiyohara:2021smr}.

On gauge configurations at nonzero temperature with the temporal extent $N_t$, the HPE of $S_{\rm eff}$ is written as
\begin{align}
    S_{\rm eff} = S_{\rm g} - N_{\rm f} N_{\rm site} \sum_{n=1}^\infty \big( \hat W(n) + \hat L(N_t,n) \big) \kappa^n ,
    \label{eq:SeffWL}
\end{align}
where $\hat W(n)$ and $\hat L(N_t,n)$ are contributions from trajectories without and with windings along the temporal direction, provided that the spatial extent is sufficiently large such that the spatial windings are negligible. We refer to the terms included in the former and the latter as the Wilson loops and the Polyakov-loop-type (PLT) loops, respectively. 
We set $\hat W(n)=\hat L(N_t,n)=0$ if corresponding closed loops of length $n$ do not exist, e.g.\ $\hat W(n)=0$ for $n<4$ or odd $n$, and $\hat L(N_t,n)=0$ for $n<N_t$.

The LO contributions from the Wilson and PLT loops are calculated to be
\begin{align}
    \hat W(4) 
    =& \ W_0(4)\, \hat P ,
    \label{eq:W(4)}
    \\
    \hat L(N_t,N_t) =& \ L_0(N_t,N_t)\, {\rm Re}\hat\Omega ,
    \label{eq:L(6,6)}
\end{align}
with the Polyakov loop
\begin{align}
\hat\Omega = \frac1{N_{\rm c}N_s^3}
\displaystyle \sum_{\vec{x}} {\rm tr_C} \left[ 
U_{\vec{x},4} U_{\vec{x}+\hat{4},4} U_{\vec{x}+2 \cdot \hat{4},4} 
\cdots U_{\vec{x}+(N_t -1) \cdot \hat{4},4} \right] ,
\label{eq:Pol}
\end{align}
where $W_0(n)$ and $L_0(N_t,n)$ are the values of $\hat{W}(n)$ and $\hat L(N_t,n)$ in the weak-coupling limit, respectively, which is obtained by setting $U_{x,\mu}=1$ for all link variables~\cite{Wakabayashi:2021eye}. The calculation of $W_0(n)$ and $L_0(N_t,n)$ is given in Appendix~\ref{sec:convergence}. For the LO we obtain
\begin{align}
    W_0(4)=96N_{\rm c}, \qquad L_0(N_t,N_t)=\frac{2^{N_t+2}N_{\rm c}}{N_t} .
\end{align}
The NLO terms are calculated to be
\begin{align}
    \hat W(6) 
    =& \ W_0(6)\, \hat P_6 ,
    \label{eq:W(6)}
    \\
    \hat L(N_t,N_t+2) 
    =& \ L_0(N_t,N_t+2)\, {\rm Re} \hat\Omega_{N_t+2},
    \label{eq:L(6,8)}
\end{align}
with $W_0(6)=2816N_{\rm c}$ and
\begin{align}
    \hat P_6 =&\ \frac1{11} \left( 3 \hat{W}_{\rm rec} + 6 \hat{W}_{\rm chair} + 2 \hat{W}_{\rm crown} \right) ,
    \label{eq:P6}
\end{align}
where $\hat{W}_{\mathrm{rec}}$, $\hat{W}_{\mathrm{chair}}$, and $\hat{W}_{\mathrm{crown}}$
represent the six-step Wilson loops of the rectangular,
chair, and crown types, respectively.
The $\hat\Omega_{N_t+2}$ in Eq.~\eqref{eq:L(6,8)} is the summation of bent PLT loops of length $N_t+2$. Definitions of these operators, as well as the derivation of these results, are found in Refs.~\cite{Kiyohara:2021smr,Wakabayashi:2021eye}.
Note that all the Wilson and PLT loops are normalized such that 
$\hat P = \hat P_6 = \hat{W}_{\mathrm{rec}} = \hat{W}_{\mathrm{chair}} = \hat{W}_{\mathrm{crown}}=1$ and 
$\hat\Omega = \hat\Omega_{N_t+2} = 1$ in the weak-coupling limit.

We rewrite the expansion~(\ref{eq:SeffWL}) as
\begin{align}
  S_{\rm eff} = S_g + S_{\rm LO} + S_{\rm NLO} + \cdots ,
  \label{eq:Sexp}
\end{align}
where $S_{\rm LO}$ and $S_{\rm NLO}$, respectively, contain the LO and NLO contributions from the Wilson and PLT loops as
\begin{align}
  S_{\rm LO} =&\ - N_{\rm f} N_{\rm site} \big( \hat W(4) \kappa^4 + \hat L(N_t,N_t) \kappa^{N_t} \big),
  \label{eq:LO}
  \\
  S_{\rm NLO} =&\ - N_{\rm f} N_{\rm site} \big( \hat W(6) \kappa^6 + \hat L(N_t,N_t+2) \kappa^{N_t+2}\big).
\end{align}
The action at the LO is given by 
\begin{align}
  S_{g+\rm LO} 
  &= S_g + S_{\rm LO}
  \nonumber \\
  &= -6 N_{\rm site} \beta^* \hat{P}
  - \lambda N_s^3 {\rm Re}\hat\Omega,
  \label{eq:g+LO}
\end{align}
with
\begin{align}
  \beta^* =& \ \beta+ \frac{N_{\rm f} W_0(4)}6\kappa^4 = \beta + 16N_{\rm f} N_{\rm c} \kappa^4 ,
  \label{eq:beta*}
  \\
  \lambda =& \ N_{\rm f} N_t L_0(N_t,N_t) \kappa^{N_t} = 2^{N_t+2} N_{\rm f} N_{\rm c} \kappa^{N_t} .
\end{align}
As shown in Ref.~\cite{Kiyohara:2021smr}, we can adopt a pseudo-heat-bath algorithm for the Monte-Carlo simulation of $S_{g+\rm LO}$ with a numerical cost comparable to that of the pure gauge theory $S_{g}$, i.e., with a cost much lower than that of the full QCD.
This enables us to carry out simulations on lattices with large spatial volumes.

\subsection{Effective incorporation of higher order terms}
\label{sec:effective}

In Ref.~\cite{Wakabayashi:2021eye}, a method to effectively incorporate the NNLO and yet higher order terms into numerical analyses has been proposed.
The key ingredient of the method is that $\hat W(n)$ and $\hat L(N_t,n)$ at different $n$, respectively, are strongly correlated with one another such that the approximate relations 
\begin{align}
    \frac{\hat W(n)}{W_0(n)} 
    \simeq&\ d_n \frac{\hat W(4)}{W_0(4)} + f_n
    = d_n \hat P + f_n, 
    \label{eq:WW=DWW}
    \\
    \frac{\hat L(N_t,n)}{L_0(N_t,n)} 
    \simeq&\ c_n \frac{\hat L(N_t,N_t)}{L_0(N_t,N_t)}
    = c_n {\rm Re}\hat\Omega,
    \label{eq:LL=CLL}
\end{align}
hold well configuration by configuration, where 
$d_n$, $f_n$, and $c_n$
are parameters common to all configurations.
In Ref.~\cite{Wakabayashi:2021eye}, the validities of Eqs.~(\ref{eq:WW=DWW}) and (\ref{eq:LL=CLL}) have been investigated on pure gauge configurations, i.e.\ in the heavy quark limit. It was found that Eq.~(\ref{eq:LL=CLL}) is well applicable up to very high $n$. Although the correlations among the Wilson loops turned out to be weaker than the PLT loops, Eq.~(\ref{eq:WW=DWW}) is also found to be well justified for small $n$. We will show later that the same tendency is also obtained on the gauge configurations generated by $S_{g+\rm LO}$ around the CP in the heavy-quark region.

Once one accepts Eqs.~(\ref{eq:WW=DWW}) and~(\ref{eq:LL=CLL}) as equalities, one can rewrite the action including the Wilson and PLT loops up to $n_{\rm W}$th and $n_{\rm L}$th orders, respectively,
\begin{align}
    & S^{(n_{\rm W},n_{\rm L})} 
    \notag \\
    &= S_{\rm g} -N_{\rm f} N_{\rm site} \Big( \sum_{n=4}^{n_{\rm W}} \hat W(n) \kappa^n + \sum_{n=N_t}^{n_{\rm L}} \hat L(N_t,n) \kappa^n \Big),
    \label{eq:SnDnL}
\end{align}
as 
\begin{align}
    S^{(n_{\rm W},n_{\rm L})} 
    \simeq S_{\rm eff\scriptscriptstyle{[LO]}}^{(n_{\rm W},n_{\rm L})} 
    = -6 N_{\rm site} \beta^*_{n_{\rm W}} \hat P -\lambda^*_{n_{\rm L}} N_s^3 {\rm Re}\hat\Omega ,
    \label{eq:SnDnL2}
\end{align}
with 
\begin{align}
    \beta^*_{n_{\rm W}}
    =&\ \beta + \frac{N_{\rm f}}6 \sum_{n=4}^{n_{\rm W}} d_n W_0(n) \kappa^n , 
    \label{eq:beta*n} 
    \\
    \lambda^*_{n_{\rm L}} 
    =& \ N_{\rm f} N_t \sum_{n=N_t}^{n_{\rm L}} c_n L_0(N_t,n) \kappa^n ,
    \label{eq:lambda*n}
\end{align}
up to an irrelevant constant.
The fact that Eq.~(\ref{eq:SnDnL2}) has the same form as Eq.~(\ref{eq:g+LO}) with the replacements $\beta^*\to\beta^*_{n_{\rm W}}$ and $\lambda\to\lambda^*_{n_{\rm L}}$ means that the numerical results at the LO can be reinterpreted as those for the action $S^{(n_{\rm W},n_{\rm L})}$ with a simple replacement of the simulation parameters. 
The measurements of higher order terms in the HPE are not necessary in this analysis once $d_n$ and $c_n$ are determined numerically.

This interpretation, however, relies on Eqs.~(\ref{eq:WW=DWW}) and (\ref{eq:LL=CLL}), which become less reliable for larger $n$.
This problem can be partially resolved if one measures the NLO terms, $\hat W(6)$ and $\hat L(N_t,N_t+2)$, on every configuration. Since the correlations of $\hat W(n)$ or $\hat L(N_t,n)$ between different $n$ are stronger for closer $n$, it is better to approximate higher order terms as 
\begin{align}
    \frac{\hat W(n)}{W_0(n)} 
    \simeq&\ \tilde d_n \frac{\hat W(6)}{W_0(6)} + \tilde f_n
    = \tilde d_n \hat P_6 + \tilde f_n,
    \label{eq:WW=DWW2}
    \\
    \frac{\hat L(N_t,n)}{L_0(N_t,n)} 
    \simeq&\ \tilde c_n \frac{\hat L(N_t,N_t+2)}{L_0(N_t,N_t+2)}
    = \tilde c_n {\rm Re}\hat\Omega_{N_t+2},
    \label{eq:LL=CLL2}
\end{align}
for $n\ge8$ and $n\ge N_t+4$, respectively,
than Eqs.~(\ref{eq:WW=DWW}) and~(\ref{eq:LL=CLL}).
Assuming Eqs.~(\ref{eq:WW=DWW2}) and~(\ref{eq:LL=CLL2}) as equalities, one obtains an approximate action of $S^{(n_{\rm W},n_{\rm L})}$ as
\begin{align}
    S_{\rm eff \scriptscriptstyle{[NLO]}}^{(n_{\rm W},n_{\rm L})}
    = S_{\rm g+LO} - N_{\rm site} \big( \gamma_{n_{\rm W}} \hat P_6 + \xi_{n_{\rm L}} {\rm Re}\hat\Omega_{N_t+2} \big),
    \label{eq:SnDnL3}
\end{align}
with 
\begin{align}
    \gamma_{n_{\rm W}} 
    =&\ N_{\rm f} \sum_{n=6}^{n_{\rm W}} \tilde d_n W_0(n) \kappa^n ,
    \\
    \xi_{n_{\rm L}} 
    =&\ N_{\rm f} \sum_{n=N_t+2}^{n_{\rm L}} \tilde c_n L_0(N_t,n) \kappa^n .
\end{align}
In later sections, we use Eq.~(\ref{eq:SnDnL3}) for the numerical analyses by performing Monte-Carlo simulations with $S_{\rm g+LO}$ and taking the effects of the second term by reweighting.
A comparison with the method using Eq.~\eqref{eq:SnDnL2}, as well as the analysis at NLO, is given in Appendix~\ref{sec:comparison}.

\begin{table}
\caption{Values of $d_n$, $c_n$, $\tilde d_n$, and $\tilde c_n$ obtained on the $36^3\times6$ lattice with $S_{g+{\rm LO}}$ near the CP at $N_{\rm f}=2$. The errors are the statistical ones.}
\label{tab:cn}
\begin{tabular}{c|cc}
\hline\hline
     $n$ &  $d_n$ & $\tilde d_n$ \\
     \hline
      4 & 1  & \\
      6 & 1.374(24) & 1 \\
      8 & 1.498(52) & 1.126(21) \\
     10 & 1.454(76) & 1.120(40) \\
     12 & 1.324(96) & 1.041(56) \\
     14 & 1.17(11)  & 0.938(68) \\
     16 & 1.03(13)  & 0.845(79) \\
     18 & 0.96(14)  & 0.797(93) \\
     20 & 1.02(18)  & 0.86(12)\\
     \hline\hline
\end{tabular}
\hspace{5mm}
\begin{tabular}{c|cc}
\hline\hline
     $n$ &  $c_n$ & $\tilde c_n$ \\ 
     \hline
      6 & 1  & \\
      8 & 0.8091(14) & 1 \\
     10 & 0.6251(16) & 0.7727(9) \\
     12 & 0.4710(15) & 0.5822(12) \\
     14 & 0.3585(13) & 0.4430(12) \\
     16 & 0.3084(14) & 0.3808(16) \\
     18 & 1.0074(67) & 1.2419(91) \\
     20 & -0.02730(6) & -0.0334(9) \\
     22 & 0.0088(3) & 0.01083(3) \\
     \hline\hline
\end{tabular}
\end{table}

\begin{figure}
  \centering
    \includegraphics[width=0.45\textwidth]{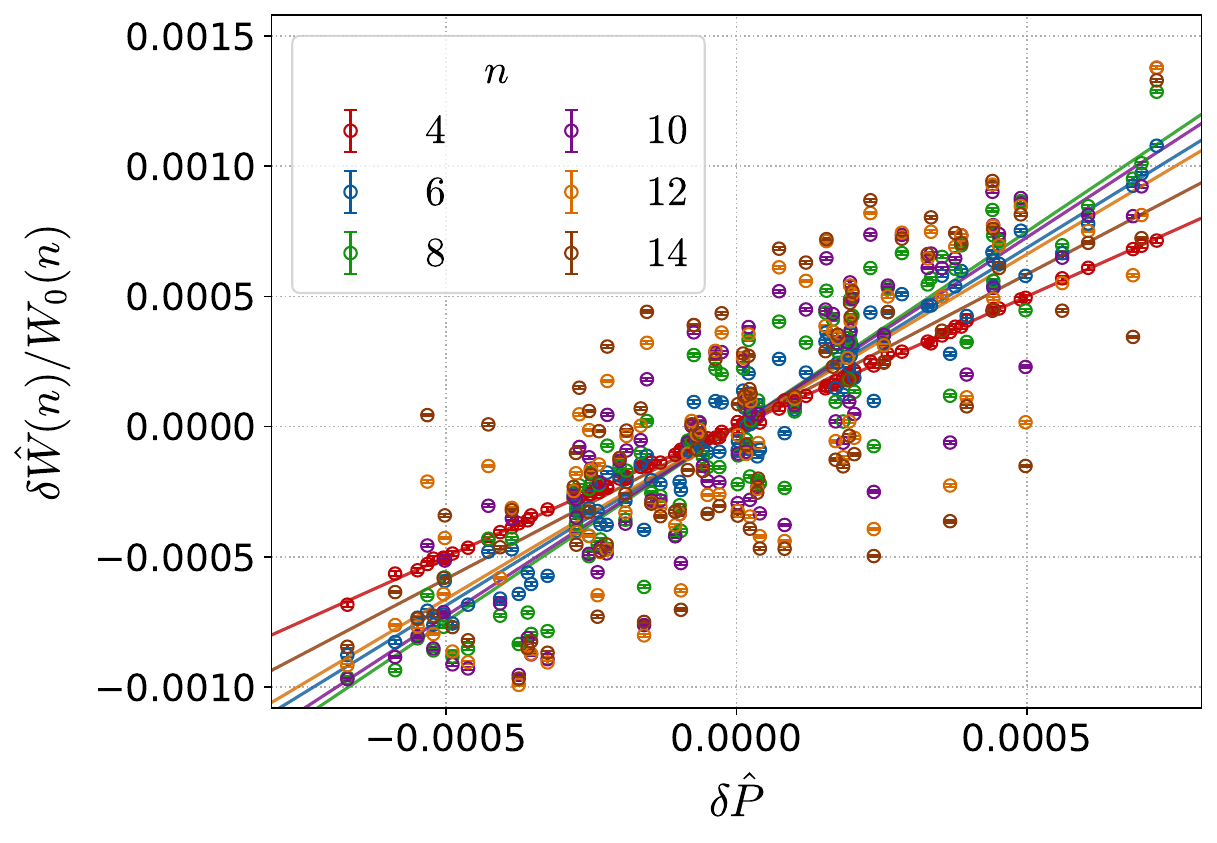}
    \includegraphics[width=0.45\textwidth]{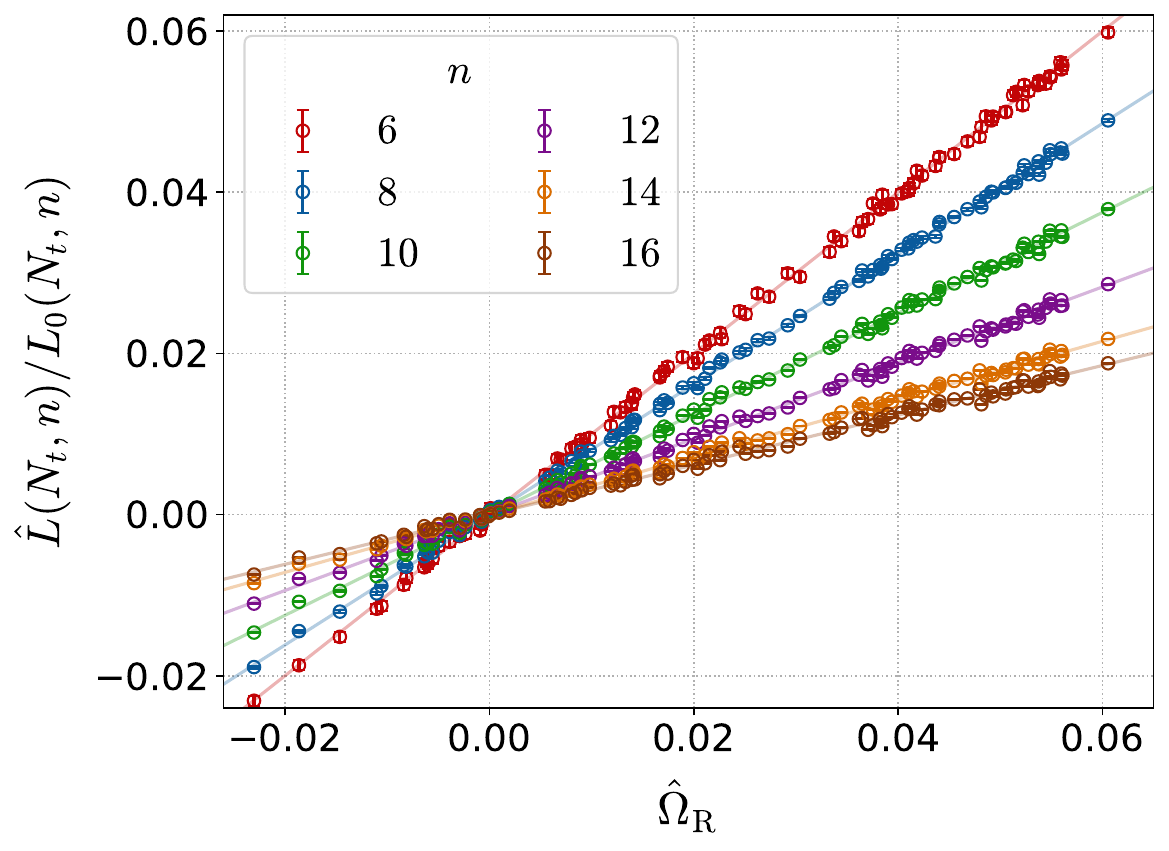}
  \caption{
    Scatter plots of $(\delta\hat W(n)/W_0(n),\delta\hat P)$ and $(\hat L(N_t,n)/L_0(N_t,n),\hat\Omega_{\rm R})$ on 100 gauge configurations of the size $N_s^3\times N_t=36^3\times6$ near the CP for various values of $n$.
    The lines show the results of the linear fits with Eqs.~\eqref{eq:WW=DWW} and \eqref{eq:LL=CLL} for each $n$.
  }
\label{fig:Lcor}
\end{figure}

\subsection{Numerical study of higher order terms}
\label{sec:DL}

Now, let us investigate the validity of the above approximate formulas for the action numerically. 
For this purpose we generate $100$ gauge configurations for the action $S_{g+\rm LO}$ with $N_{\rm f}=2$ on the $36^3\times6$ lattice at the parameters close to the CP at the LO, ($\beta^*,\lambda)=(5.8905,0.0012)$. We measure $\hat W(n)$ and $\hat L(N_t,n)$ on each configuration through the measurement of $B^n$ by the noise method~\cite{Wakabayashi:2021eye} with $5,000$ noise vectors.

\begin{figure*}
  \centering
    \includegraphics[width=0.45\textwidth]{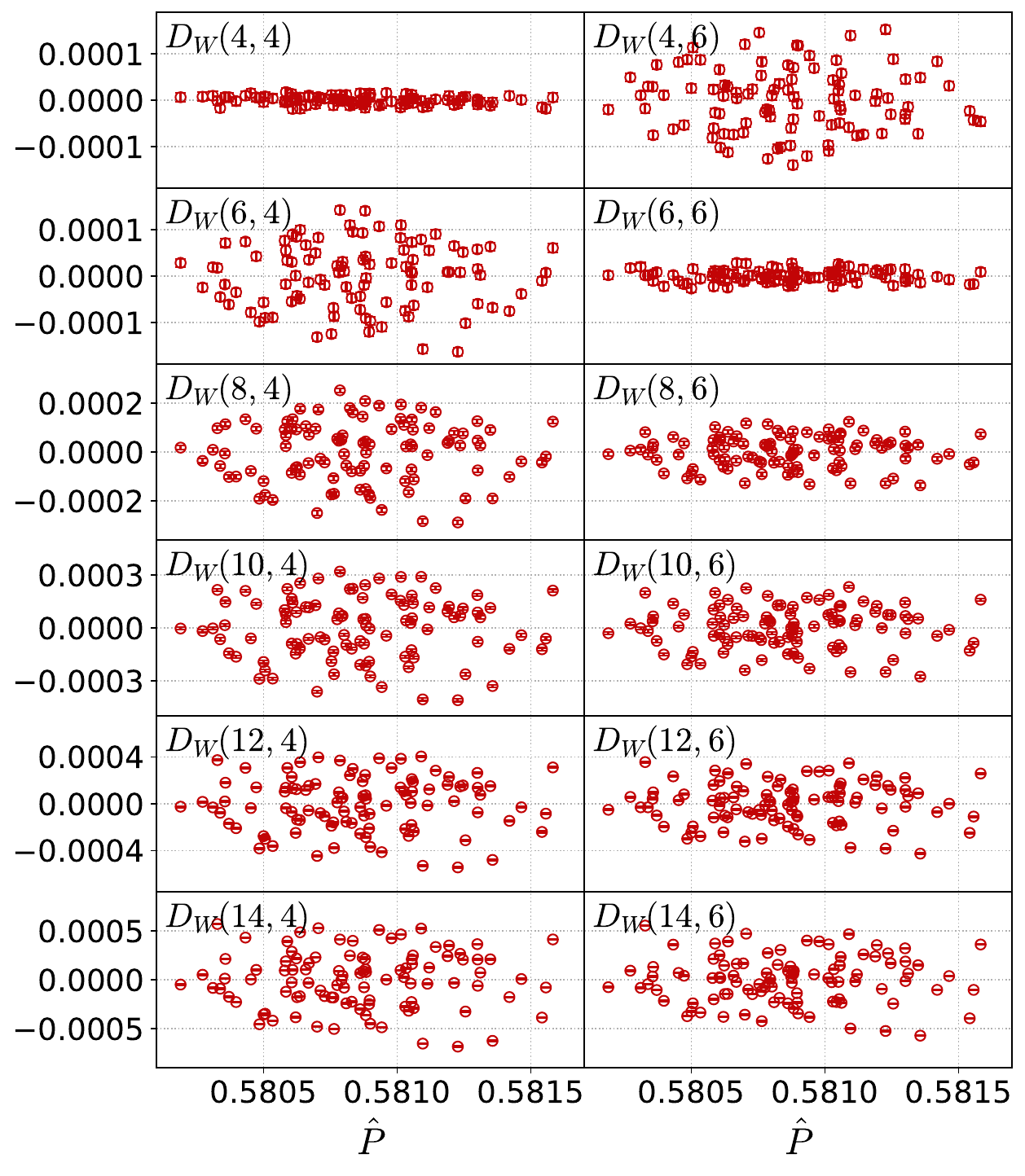}
    \includegraphics[width=0.45\textwidth]{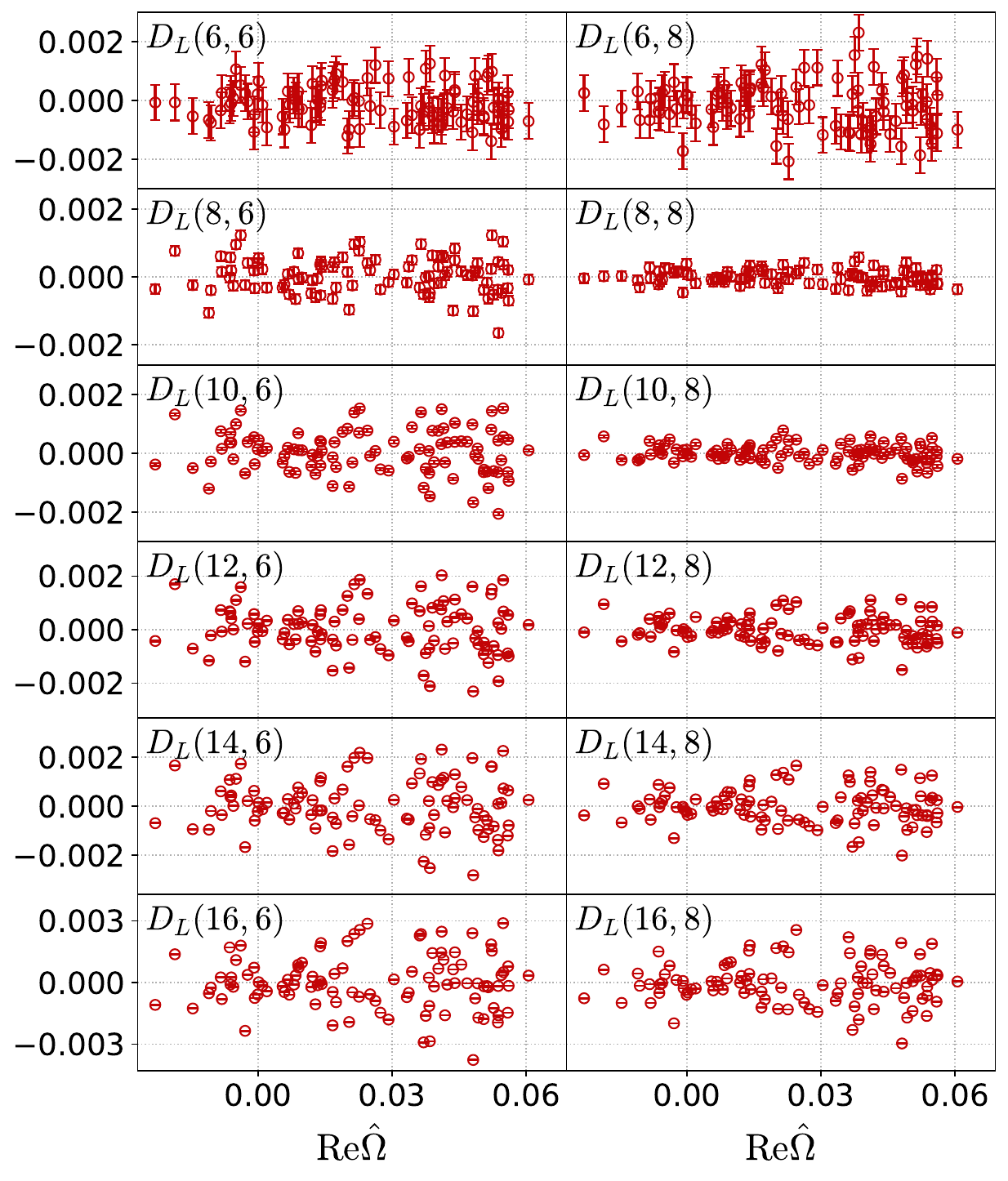}
  \caption{
  Scatter plots of $({\hat D}_W(n,m),\delta\hat P)$ and $({\hat D}_L(n,m),\hat\Omega_{\rm R})$ for various values of $n$ and $m$. The values measured by the noise method are used for $\hat W(n)$ and $\hat L(N_t,n)$ in Eqs.~\eqref{eq:DWnm} and~\eqref{eq:DLnm}, while $\hat W(m)$ and $\hat L(N_t,m)$ are exactly measured values. 
  Nonzero scatterings in $\hat D_W(n,n)$ and $\hat D_L(n,n)$ come from the noise method.
  }
\label{fig:Lcor2}
\end{figure*}

In Fig.~\ref{fig:Lcor}, we show the scatter plots of $(\delta\hat W(n)/W_0(n), \hat P)$ and $(\hat L(N_t,n)/L_0(N_t,n), \hat\Omega_{\rm R}={\rm Re}\hat\Omega)$ measured on every configuration for various $n$ with $\delta\hat W(n) = \hat W(n)-\langle\hat W(n)\rangle$. 
The figure shows that each result distributes linearly, indicating the strong correlations~(\ref{eq:WW=DWW}) and (\ref{eq:LL=CLL}), whereas the distribution tends to be more scattered as $n$ becomes larger, especially in $\hat W(n)$.

The coefficients $d_n$, $c_n$, $\tilde d_n$, and $\tilde c_n$ determined from these results by the linear fits are summarized in Table~\ref{tab:cn}, where the statistical errors are estimated by the jackknife method. The errors from the noise method are well suppressed compared to the statistical ones.\footnote{As discussed in Ref.~\cite{Wakabayashi:2021eye}, concerning $c_n$ and $\tilde c_n$ in Table~\ref{tab:cn}, the cases $n=18$ and $20$ for $N_t=6$ are special because the sign of $L_0(N_t,n)$ changes between $n=18$ and $20$ and thus $L_0(N_t,n)$ suffers from strong cancellation of positive and negative terms at $n=18$ and $20$. For $N_t=8$, this happens first between $n=26$ and $28$.}
The results of $d_n$ and $c_n$ are close to those obtained on pure gauge simulations~\cite{Wakabayashi:2021eye}, suggesting a mild dependence of $d_n$ and $c_n$ on $\lambda$ and $\beta^*$ near the phase transition. We thus employ the values given in Table~\ref{tab:cn} in the following analyses.

To see the correlations in more detail, in Fig.~\ref{fig:Lcor2} we plot
\begin{align}
    \hat D_W(n,m) =&\ \frac{\delta \hat W(n)}{d_n W_0(n)} - \frac{\delta \hat W(m)}{d_m W_0(m)} ,
    \label{eq:DWnm}
    \\
    \hat D_L(n,m) =&\ \frac{\hat L(N_t,n)}{c_n L_0(N_t,n)} - \frac{\hat L(N_t,m)}{c_m L_0(N_t,m)} .
    \label{eq:DLnm}
\end{align}
These quantities vanish when the $n$th- and $m$th-order terms are perfectly correlated. Since we measure exact values of $\hat W(4)$, $\hat W(6)$, $\hat L(6,6)$, $\hat L(6,8)$ in our simulations, we use them for $\hat W(m)$ and $\hat L(N_t,m)$, while the values measured by the noise method are used for $\hat W(n)$ and $\hat L(N_t,n)$ in Eqs.~\eqref{eq:DWnm} and~\eqref{eq:DLnm}. The figure shows that their distribution around zero becomes broader as $n$ becomes larger with fixed $m$. We also note that the scattering is more suppressed in $\hat D_W(n,6)$ ($\hat D_L(n,8)$) than $\hat D_W(n,4)$ ($\hat D_L(n,6)$) for $n\ge8$ ($n\ge10$). This result shows that Eqs.~(\ref{eq:WW=DWW2}) and (\ref{eq:LL=CLL2}) are superior to Eqs.~(\ref{eq:WW=DWW}) and (\ref{eq:LL=CLL}), respectively, as expected. One also sees that the correlations between the PLT loops survive up to large $n$, while those between the Wilson loops tend to be blurred more quickly~\cite{Wakabayashi:2021eye}.

\begin{figure}
  \centering
    \includegraphics[width=0.45\textwidth]{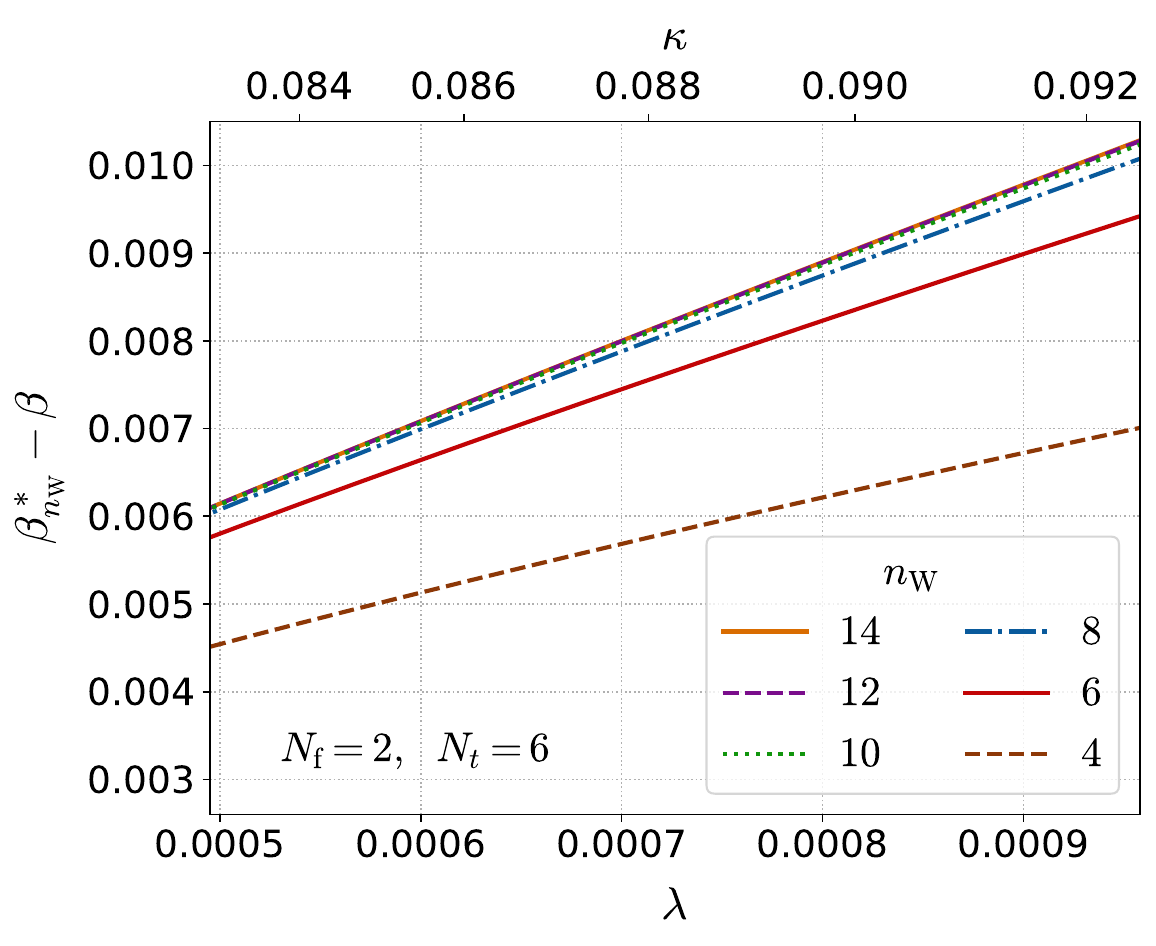}
    \includegraphics[width=0.45\textwidth]{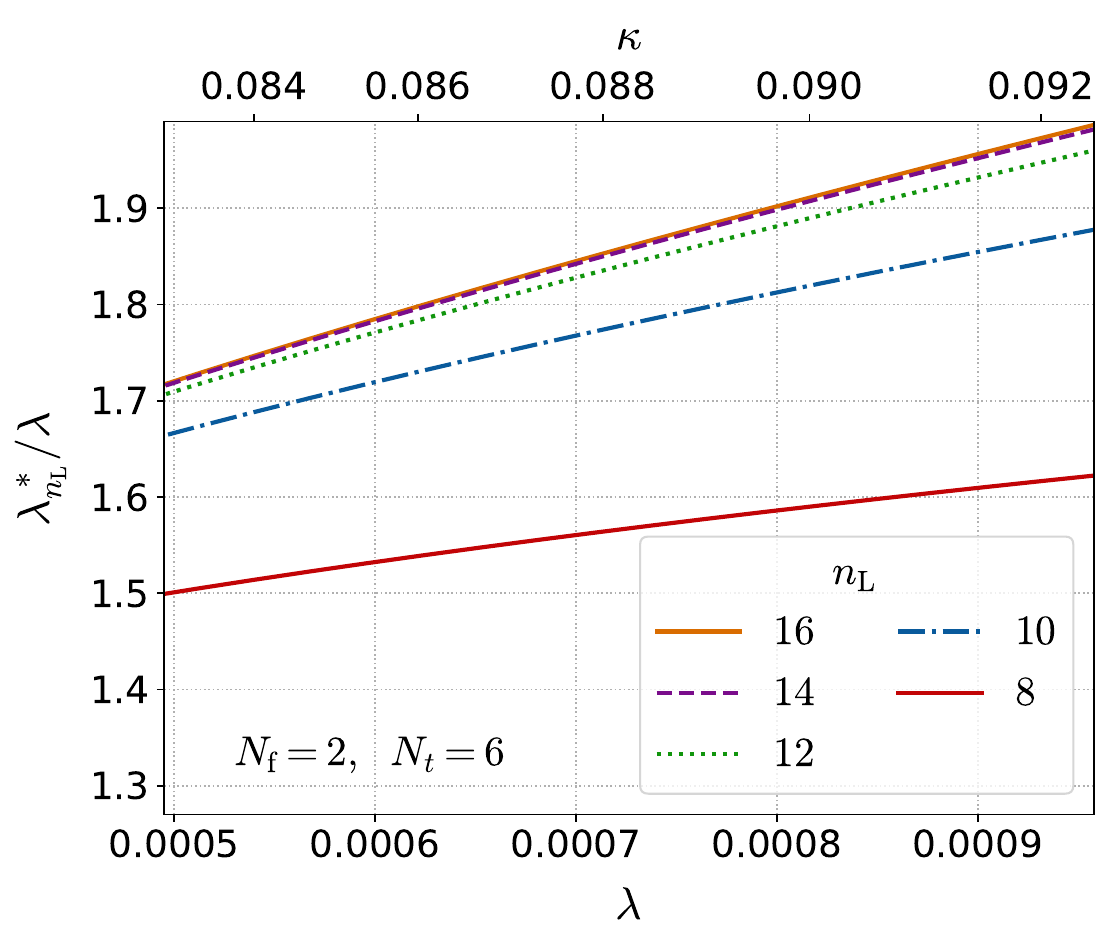}
  \caption{
  $\lambda$ dependence of $\beta^*_{n_{\rm W}}$ and $\lambda^*_{n_{\rm L}}$ for various $n_{\rm W}$ and $n_{\rm L}$ for $N_{\rm f}=2$. The top scale shows the value of $\kappa$.
  }
\label{fig:star}
\end{figure}

In Fig.~\ref{fig:star}, we plot the coefficients in Eq.~(\ref{eq:SnDnL2}), $\beta^*_n$, $\lambda^*_n$,
as functions of $\lambda$ in the range relevant for later analyses for various $n$; as we will see in Sec.~\ref{sec:result}, the CP is located at $\lambda\simeq0.0007$. The upper and lower panels plots $\beta^*_n-\beta$ and $\lambda^*_n/\lambda$. Notice that these quantities represent the magnitude of the quark contributions to the action at the $n$th order provided the validity of Eqs.~(\ref{eq:WW=DWW}) and (\ref{eq:LL=CLL}).
The figure shows that the coefficients almost converge around $n_{\rm W}=10$ and $n_{\rm L}=14$, indicating the convergence of the HPE. These results agree with the case of free fermions discussed in Appendix~\ref{sec:free}.

Applicability of Eq.~(\ref{eq:SnDnL3}) may become less justified with large $n_{\rm W}$ and $n_{\rm L}$. 
However, the rapid convergence of the HPE shown in Fig.~\ref{fig:star} indicates that the contribution of the higher-order terms is more suppressed. 
From Fig~\ref{fig:Lcor2}, the correlations~(\ref{eq:WW=DWW2}) and (\ref{eq:LL=CLL2}) hold well until $n_{\rm W}=10$ and $n_{\rm L}=14$ where the contribution of higher order terms is well negligible in Fig.~\ref{fig:star}. Therefore, to a good approximation Eq.~(\ref{eq:SnDnL3}) can be regarded as the action that includes all orders of the HPE near the CP in the heavy-quark region.
In the following, we employ Eq.~(\ref{eq:SnDnL3}) with 
\begin{align}
    (n_{\rm W},n_{\rm L}) = (10,14) ,
\end{align}
for numerical analyses, whereas the use of yet larger $n_{\rm W}$ and $n_{\rm L}$ hardly changes the following results.

\section{Numerical setup for Binder-cumulant analysis}
\label{sec:detail}

We generate gauge configurations with the action $S_{g+\rm LO}$ using the method proposed in Ref.~\cite{Kiyohara:2021smr}. 
In our previous study at $N_t=4$, it was found that this method enables high-precision measurements of observables around the CP in the heavy-quark region thanks to efficient Monte-Carlo updates avoiding the overlapping problem in reweighting~\cite{Kiyohara:2021smr}. 
As we will see below, the method works well also at $N_t=6$.

In the present study, we perform the numerical analyses with $N_t=6$, while the 
spatial lattice size $N_s^3$ is varied in the range of the aspect ratio $6\le LT=N_s/N_t \le18$.
For each $LT$, gauge configurations are generated for $S_{g+\rm LO}$ at $4$--$6$ sets of $(\beta^*,\lambda)$ given in Table~\ref{tab:params}.
The value of $\beta^*$ for each $\lambda$ is chosen to be close to the transition point at the LO.

The gauge configurations are updated by the pseudo-heat-bath (PHB) and over-relaxation (OR) algorithms~\cite{Kiyohara:2021smr}.
Each PHB step is followed by five OR steps. We measure observables for every $5$ set of the PHB+OR updates for $LT\le10$. The separation is enlarged to $20$ ($40$) sets for $LT=12,15$ ($LT=18$).
We have performed $10^6$ measurements at each $(\beta^*,\lambda)$ for $8\le LT \le 12$, while the number of measurements is $5\cdot10^5$ for $LT=6,7$ and $15\cdot10^5$ ($8\cdot10^5$) for $LT=15$ ($LT=18$).

\begin{table}[bt]
  \centering
  \caption{Simulation parameters $\beta^*$ and $\lambda$. The right columns denote the values of $LT$ at which the simulations are performed at the parameters in the left columns.}
  \label{tab:params}
  \begin{tabular}{ll|cccccccc}
    \hline
    & & \multicolumn{8}{l}{$LT$} \\
    $\lambda$ & $\beta^*$ & $6$ & $7$ & $8$ & $9$ & $10$ & $12$ & $15$ & $18$ \\
    \hline
    $0.0008$ & $5.8918$ & $\circ$ & $\circ$ & $\circ$ & $\circ$ & $\circ$ \\
    $0.0010$ & $5.8911$ & $\circ$ & $\circ$ & $\circ$ & $\circ$ & $\circ$ & $\circ$ & $\circ$ \\
    $0.0012$ & $5.8905$ & $\circ$ & $\circ$ & $\circ$ & $\circ$ & $\circ$ & $\circ$ & $\circ$ & $\circ$ \\
    $0.0013$ & $5.8901$ & &&&&&&& $\circ$ \\
    $0.0014$ & $5.8899$ & $\circ$ & $\circ$ & $\circ$ & $\circ$ & $\circ$ & $\circ$ & $\circ$ \\
    $0.0014$ & $5.8897$ & &&&&&&& $\circ$ \\
    $0.0015$ & $5.8894$ & &&&&&&& $\circ$ \\
    $0.0016$ & $5.8892$ & $\circ$ & $\circ$ & $\circ$ & $\circ$ & $\circ$ & $\circ$ & $\circ$ \\
    $0.0018$ & $5.8886$ & $\circ$ & $\circ$ & $\circ$ & $\circ$ & $\circ$ & $\circ$ \\
    \hline
  \end{tabular}
\end{table}

\begin{figure}
  \centering
    \includegraphics[width=0.45\textwidth]{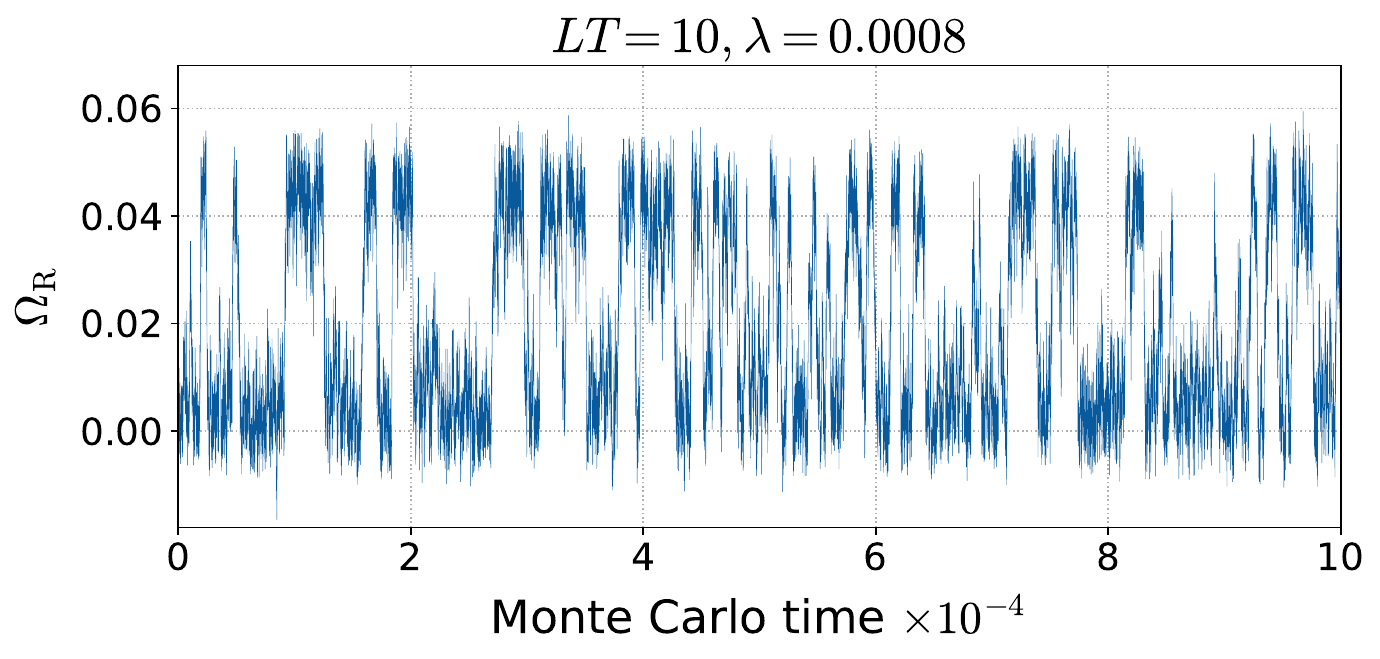}
    \includegraphics[width=0.45\textwidth]{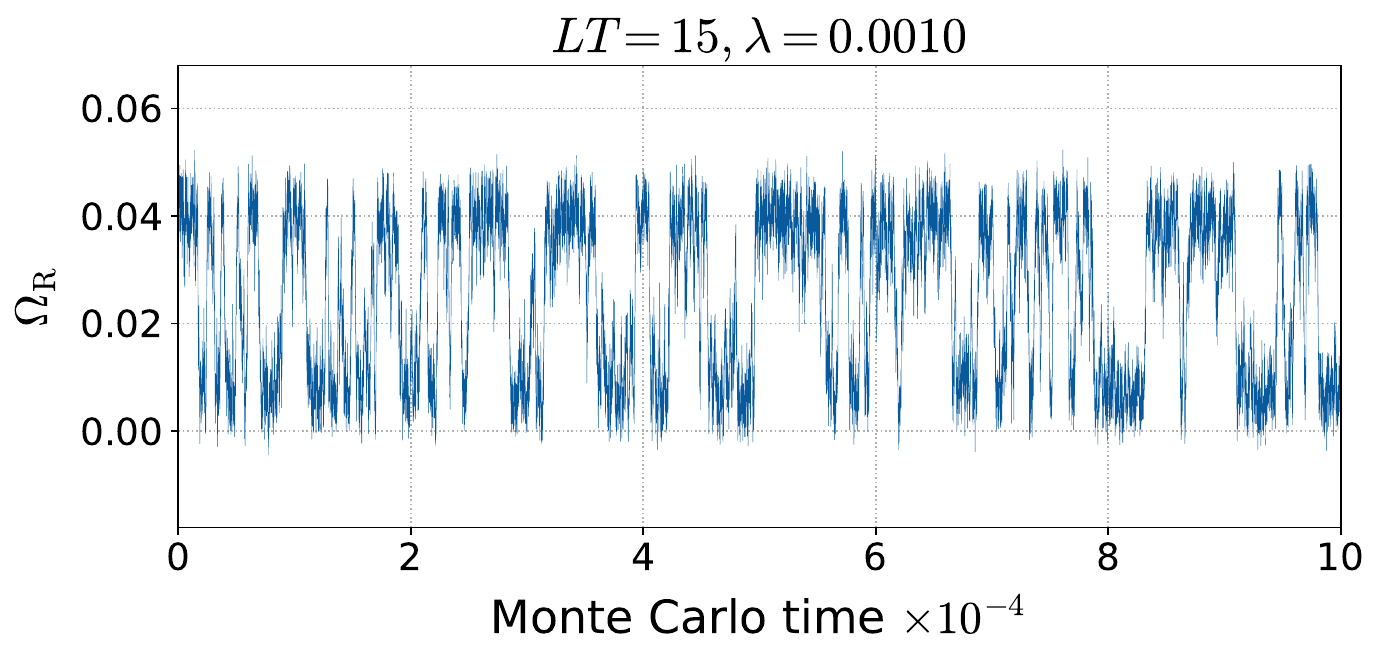}
    \includegraphics[width=0.45\textwidth]{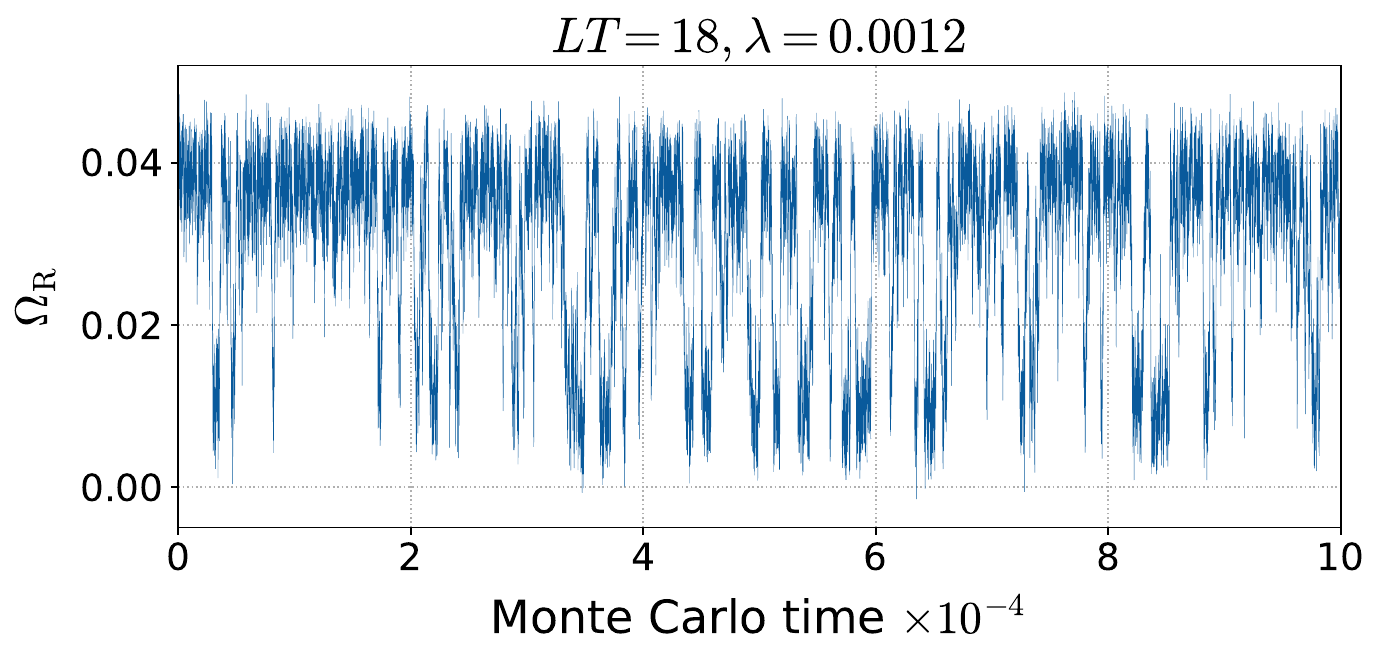}
  \caption{
    Monte-Carlo time history of $\hat\Omega_{\rm R}$ for $LT=10,15,18$ at the smallest $\lambda$ in the simulations. The horizontal axis is the Monte-Carlo time in the unit of measurement. The time history for $10^5$ measurements is shown for visibility.
  }
\label{fig:history}
\end{figure}

Near a first-order phase transition, the autocorrelation length in the Monte-Carlo time tends to be large due to the suppression of the flippings between meta-stable states. This problem becomes more serious on larger lattices, as well as for smaller $\lambda$ where the first-order transition is stronger. To see the rate of flippings in the Monte-Carlo time, we show in Fig.~\ref{fig:history} the Monte-Carlo history of $\hat\Omega_{\rm R}$ at the smallest $\lambda$ for $LT=10$ (top), $15$ (middle), and $18$ (bottom). The horizontal axis is the Monte-Carlo time in the unit of measurement and the figure shows $10^5$ measurements among the total ones for visibility. 
From the figure, one sees that the flipping occurs frequently within the Monte-Carlo time shown in the figure. The autocorrelation length corresponding to each result is less than $2,000$ in the unit of measurement. 
The autocorrelation length becomes shorter for larger $\lambda$ with fixed $LT$.

To estimate the statistical errors of observables, we use the jackknife analysis with the bin size $25,000$, unless otherwise stated. This bin size is significantly larger than the autocorrelation length and large enough to eliminate its effect. In fact, we have checked that statistical errors do not have a clear dependence on the bin size within the interval $10,000$--$40,000$.

On each set of simulation parameters, $(\beta^*,\lambda)$, we measure $\hat P$, $\hat P_6$, $\hat\Omega$, $\hat\Omega_{N_t+2}$, and other observables on every measurement. 
From these results, we calculate observables at a physical parameter $(\beta_{\rm ph},\lambda_{\rm ph})$ for the action~\eqref{eq:SnDnL3} by reweighting, where both the effects of the parameter shift $(\beta^*,\lambda)\to(\beta_{\rm ph},\lambda_{\rm ph})$ and the higher-order terms in the HPE are taken into account. The numerical results thus obtained from individual simulation parameters are averaged over to obtain the final result for each $LT$.

\section{Binder cumulant}
\label{sec:result}

In this section, we perform the numerical analysis of the Binder cumulant of $\hat\Omega_{\rm R}$ near the CP of heavy-quark QCD. 
In the following, we set $N_{\rm f}=2$ unless otherwise stated.

\subsection{FSS of Binder cumulant}
\label{sec:FSS}

The Binder cumulant~\cite{Binder:1981sa} of $\hat\Omega_{\rm R}$ is defined by 
\begin{align}
    B_4 = \frac{\langle \hat\Omega_{\rm R}^4 \rangle_{\rm c}+3}{\langle \hat\Omega_{\rm R}^2 \rangle_{\rm c}^2} ,
    \label{eq:B4}
\end{align}
where $\langle \hat\Omega_{\rm R}^n \rangle_{\rm c}$ is the $n$th-order cumulant of $\hat\Omega_{\rm R}$.

The CP in heavy-quark QCD is believed to belong to the $Z(2)$ universality class, to which the CP in the three-dimensional Ising model also belongs. 
If $\hat\Omega_{\rm R}$ corresponds to the magnetization in the Ising model, from the FSS argument it is shown that the minimum of $B_4$ with the variation of $\beta$ with fixed $\lambda$ behaves as 
\begin{align}
    B_4(\lambda,LT) = b_4 + c( \lambda-\lambda_{\rm c} ) (LT)^{1/\nu} ,
    \label{eq:B4ansatz4}
\end{align}
near the CP, where the higher-order terms of $\lambda-\lambda_{\rm c}$ and $1/LT$ are neglected. The CP exists at $\lambda=\lambda_{\rm c}$, at which $B_4(\lambda,LT)$ with different $LT$ cross at the universal value $b_4$. In the $Z(2)$ universality class, the values of $b_4$ and $\nu$ are~\cite{Pelissetto:2000ek}
\begin{align}
b_4=1.604, \qquad \nu=0.630,
\label{eq:b4nu}
\end{align}
whereas the values of $\lambda_{\rm c}$ and $c$ are not constrained by the universality.

When $\hat\Omega_{\rm R}$ is not solely given by the magnetization but corresponds to a linear combination of the magnetic- and energy-like observables in the Ising model, Eq.~\eqref{eq:b4nu} is modified as~\cite{Jin:2013wta,Cuteri:2020yke}
\begin{align}
    B_4(\lambda,LT) 
    = \big( b_4 + c( \lambda-\lambda_c ) (LT)^{1/\nu} \big) \big( 1 + d(LT)^Y\big),
    \label{eq:B4ansatz6}
\end{align}
with
\begin{align}
    Y=
    -0.894 ,
    \label{eq:Y}
\end{align}
in the $Z(2)$ universality class~\cite{Pelissetto:2000ek}.
In this case, $B_4(\lambda,LT)$ with different $LT$ no longer cross at a point.

\begin{figure*}
  \centering
    \includegraphics[width=0.32\textwidth]{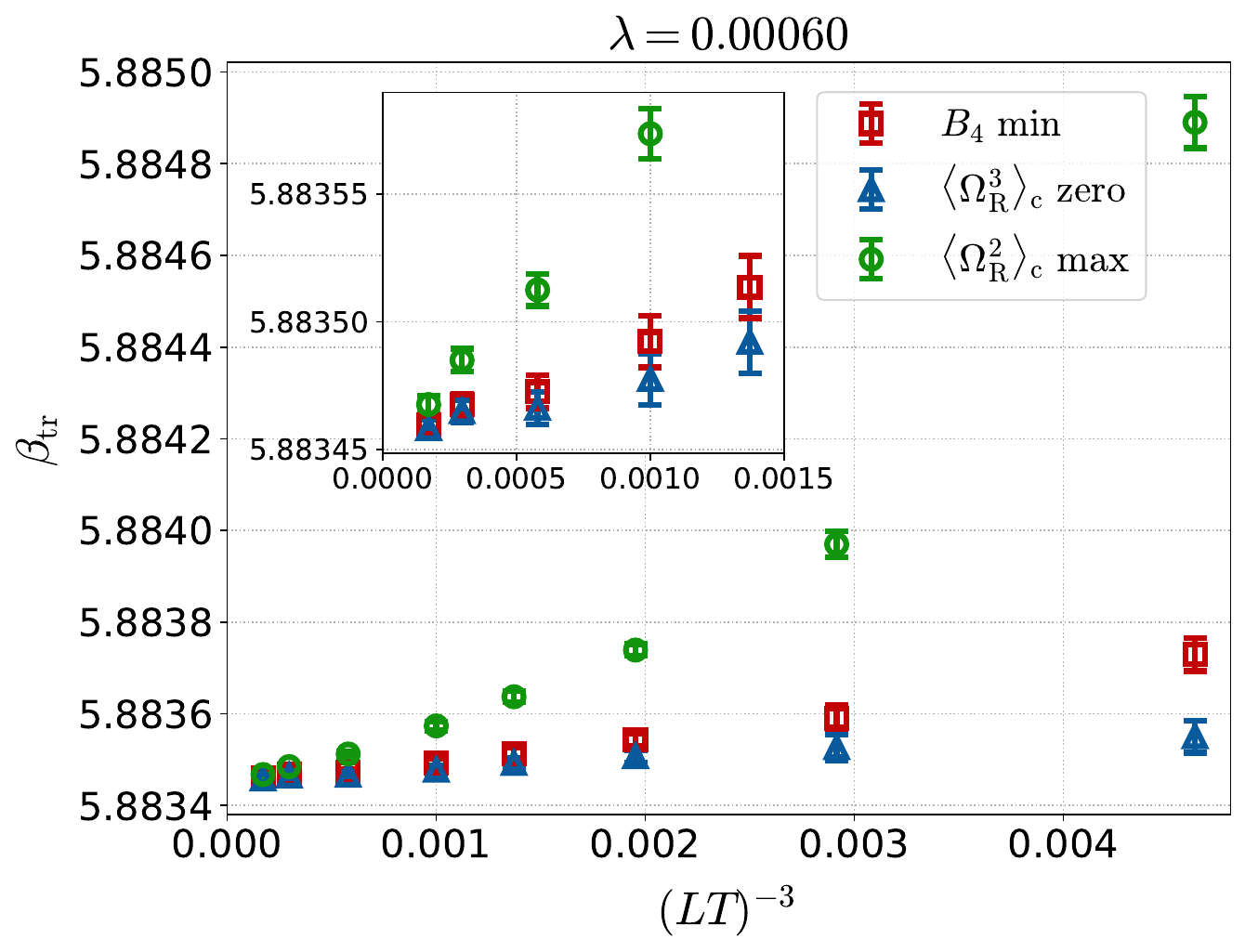}
    \includegraphics[width=0.32\textwidth]{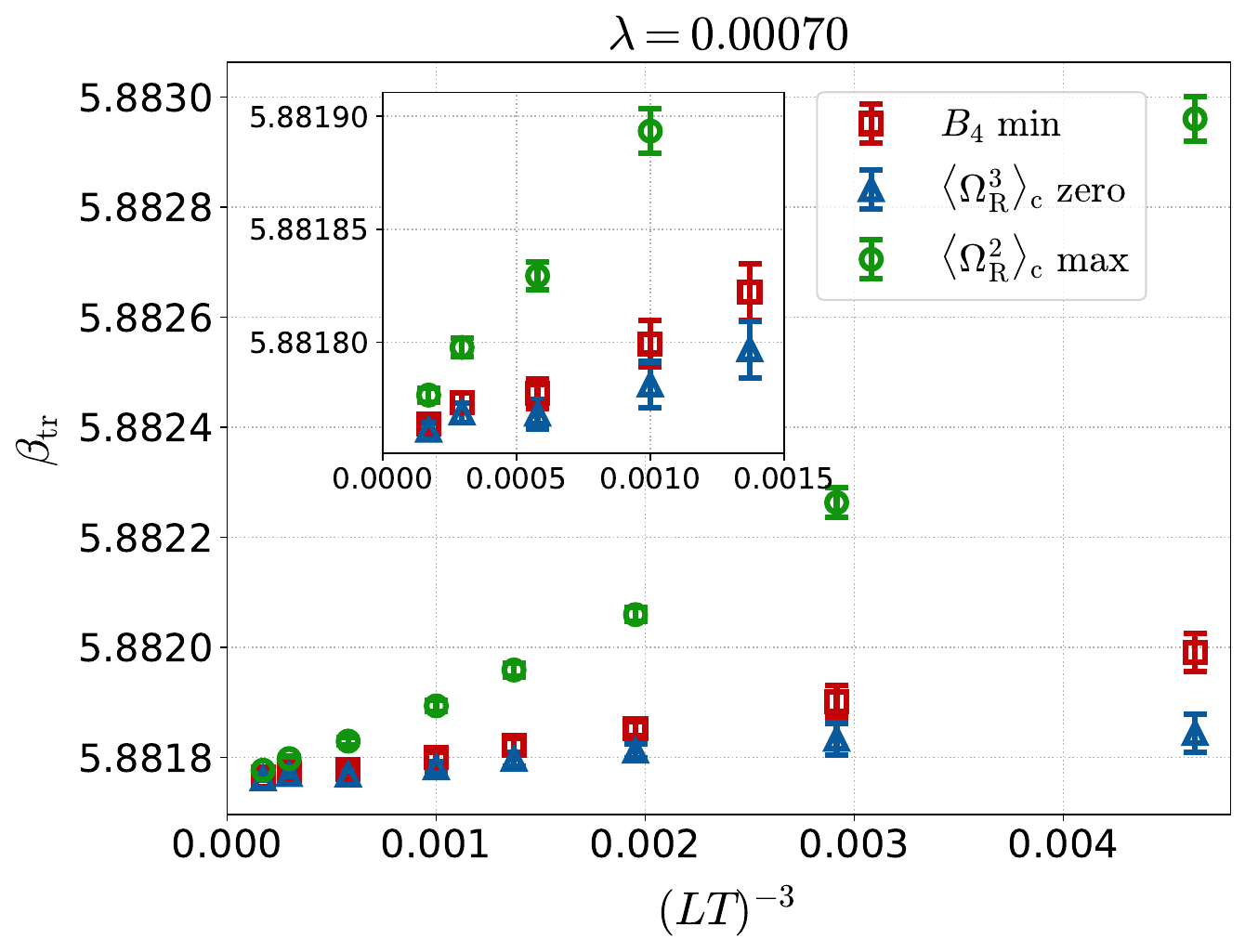}
    \includegraphics[width=0.32\textwidth]{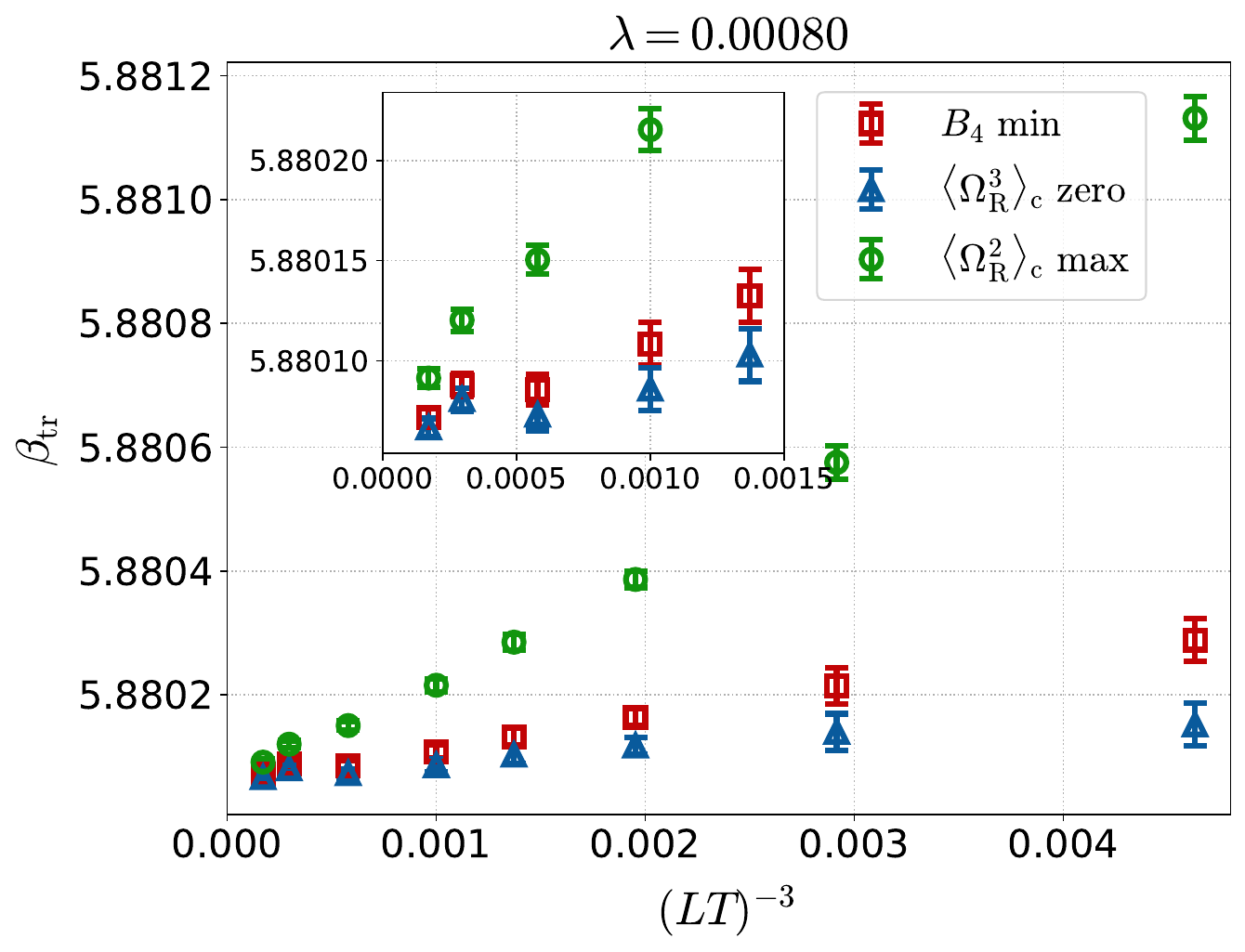}
  \caption{
    Transition point $\beta_{\rm tr}(\lambda)$ determined by (i) minimum of $B_4$, (ii) zero of $\langle\Omega_{\rm R}^3\rangle_{\rm c}$, and (iii) maximum of $\langle \Omega_{\rm R}^2\rangle_{\rm c}$, as a function of $(LT)^{-3}$ for $\lambda=0.0006$, $0.0007$, and $0.0008$.
  }
\label{fig:transitionLT}
\end{figure*}

\begin{figure}
  \centering    \includegraphics[width=0.4\textwidth]{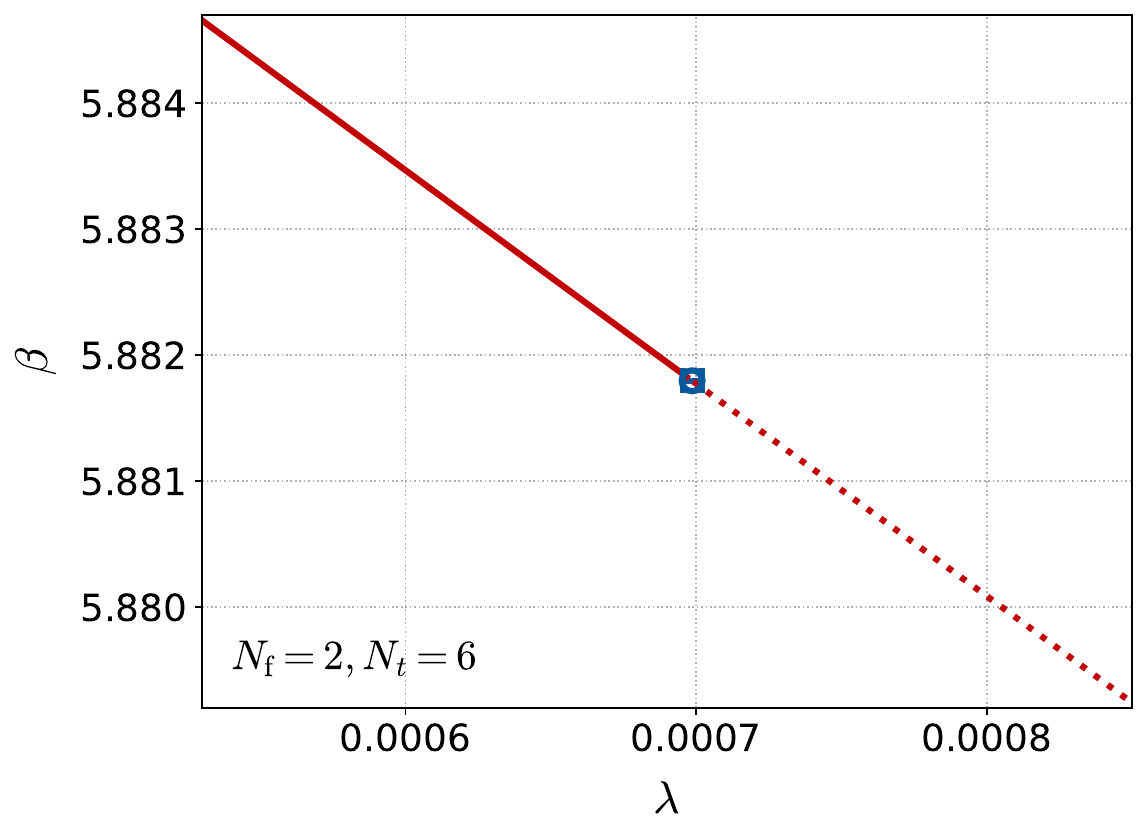}
  \includegraphics[width=0.4\textwidth]{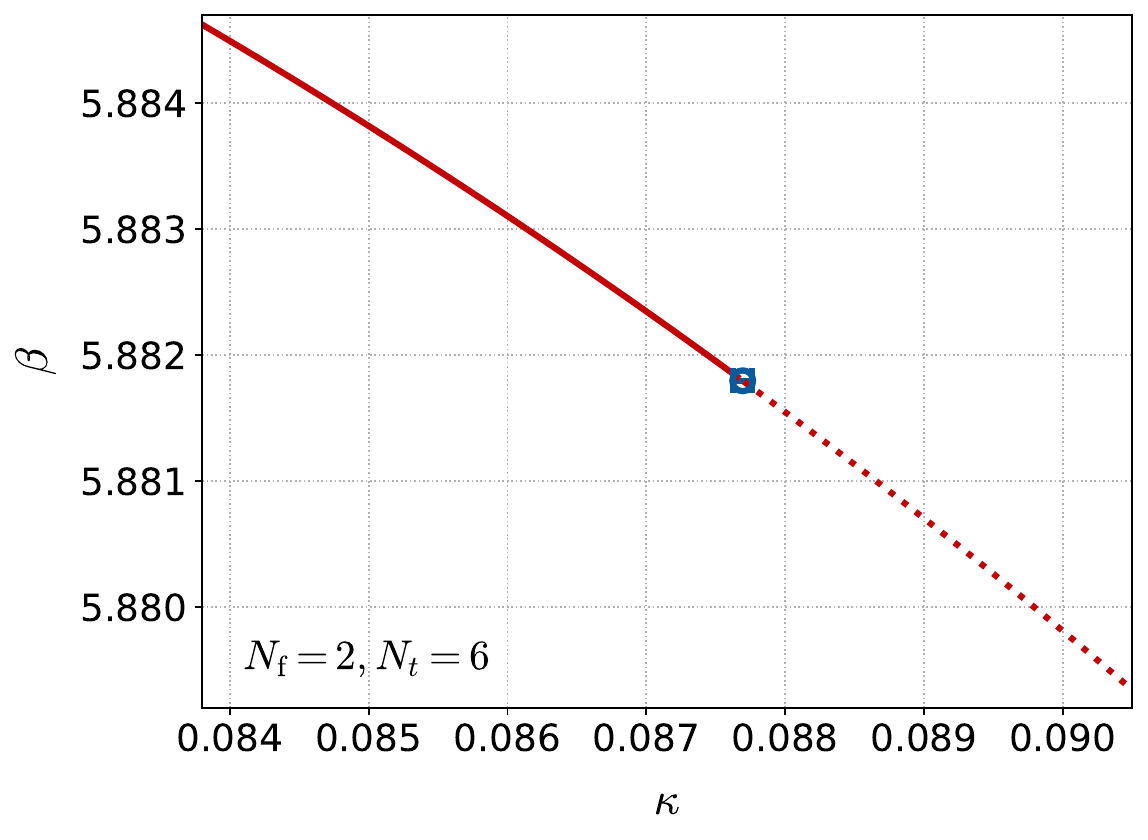}
  \caption{
    Transition line $\beta_{\rm tr}(\lambda)$ as a function of $\lambda$ (upper) and $\kappa$ (lower) obtained from the minimum of $B_4$ at $LT=15$. The CP determined by the fit with Eq.~\eqref{eq:B4ansatz6} is shown by the circle marker. The solid and dashed lines show the first-order transition and crossover. 
    Errors of the line are smaller than the thickness.
  }
\label{fig:transition}
\end{figure}

\subsection{Numerical results}
\label{sec:numerical}

The use of the minimum value of $B_4$ in the above argument is motivated by the observation that $B_4$ takes the minimum at the transition point $\beta=\beta_{\rm tr}(\lambda)$, which corresponds to the vanishing external magnetic field in the Ising model, for a given $\lambda$. 
Alternatively, $\beta_{\rm tr}(\lambda)$ may also be defined through~\cite{Kiyohara:2021smr}
\begin{itemize}
    \item Maximum of $\langle\Omega_{\rm R}^2\rangle_{\rm c}$,
    \item Zero of $\langle\Omega_{\rm R}^3\rangle_{\rm c}$.
\end{itemize}
To check the consistency of these three definitions of $\beta_{\rm tr}(\lambda)$, 
in Fig.~\ref{fig:transitionLT} we plot $\beta_{\rm tr}(\lambda)$ obtained by each definition for $\lambda=0.0006$ (left), $0.0007$ (middle), and $0.0008$ (right) as functions of $(LT)^{-3}$. The figure shows that all the definitions for $\beta_{\rm tr}(\lambda)$ converge to the same value in the $LT\to\infty$ limit. Among them, the definition through $\langle\Omega_{\rm R}^2\rangle_{\rm c}$ has the strongest dependence on $LT$, while the dependence is milder in the other definitions. 
In Fig.~\ref{fig:transition}, we show $\beta_{\rm tr}(\lambda)$ obtained from the minimum of $B_4$ at $LT=15$ on the $\beta$--$\lambda$ and $\beta$--$\kappa$ planes. The statistical errors are smaller than the thickness of the line.

\begin{figure}
  \centering
    \includegraphics[width=0.4\textwidth]{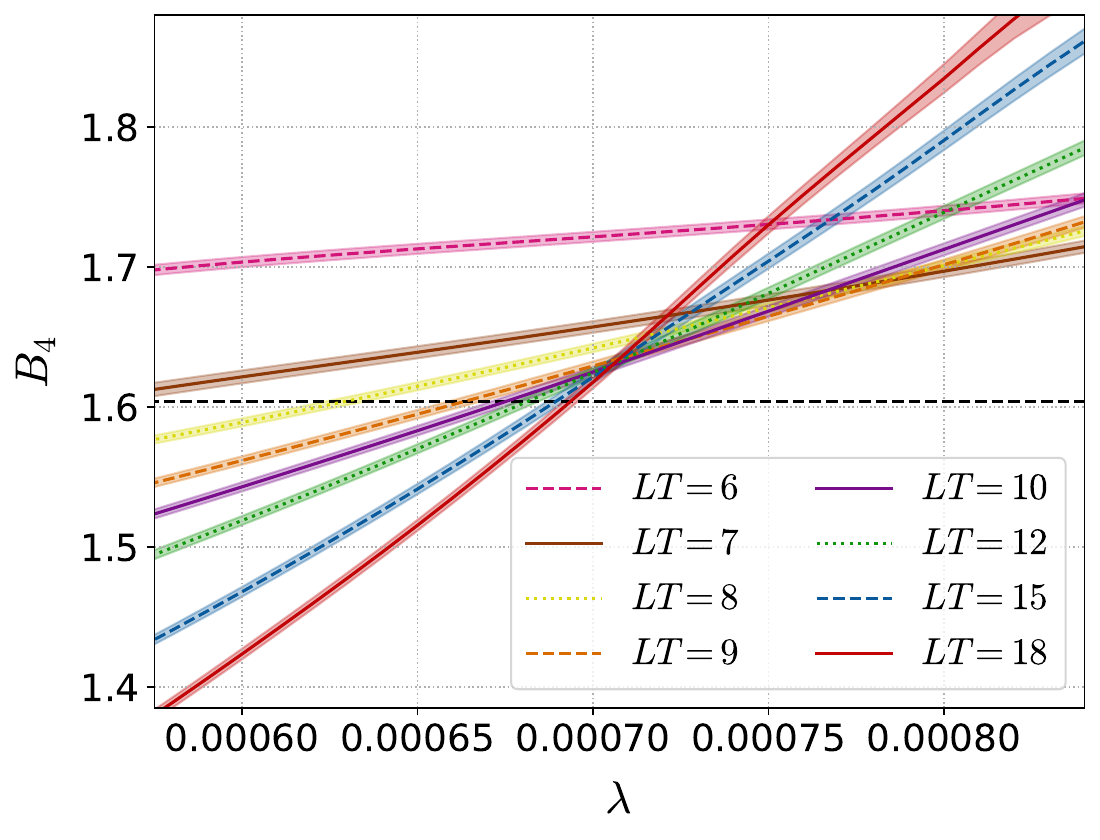}
    \includegraphics[width=0.4\textwidth]{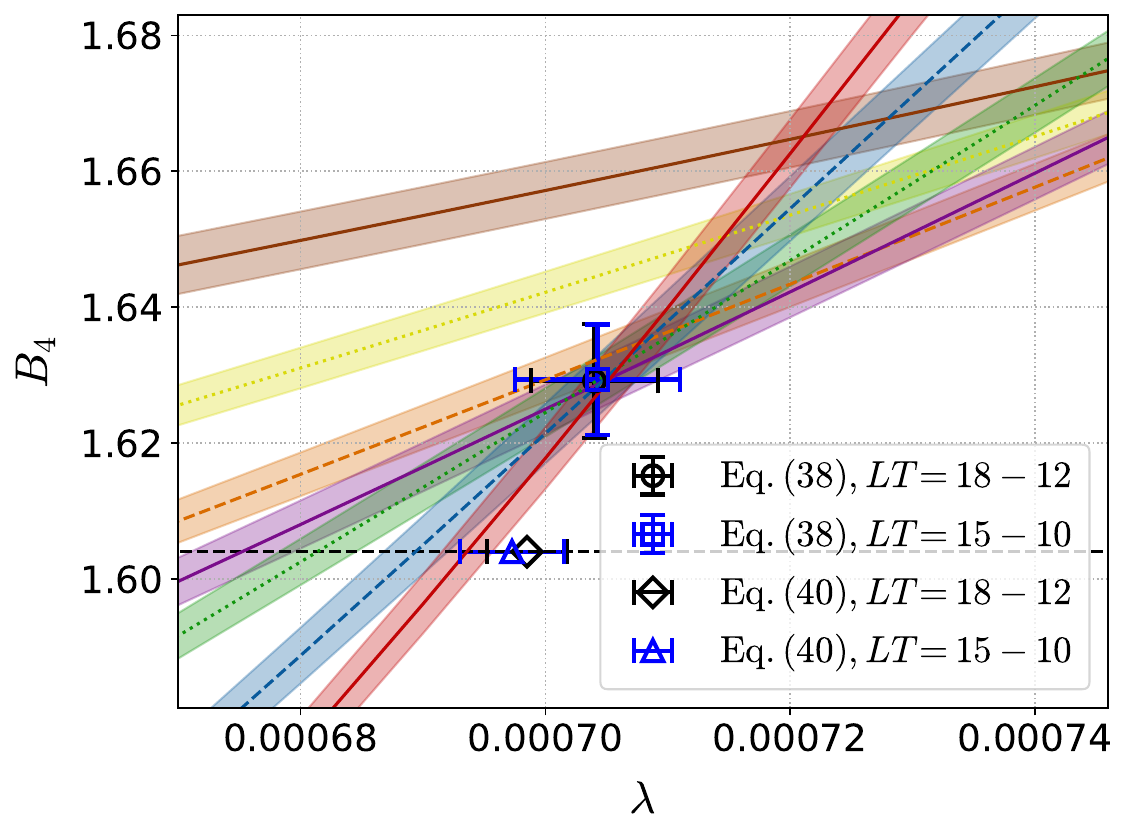}
  \caption{
    $\lambda$ dependence of the minimum of $B_4$ obtained at various $LT$. The lower panel is an enlargement of the upper around the crossing point. The circle (square) symbol shows the fit result of the crossing point based on Eq.~(\ref{eq:B4ansatz4}) with $LT\ge12$ ($LT\ge10$). The diamond and triangle symbols at $B_4=b_4$ denote the fit results based on Eq.~\eqref{eq:B4ansatz6}. 
  }
\label{fig:B4}
\end{figure}

In Fig.~\ref{fig:B4}, we show the $\lambda$ dependence of the minimum of $B_4$ obtained on various $LT$. The lower panel is an enlargement of the upper. Statistical errors are shown by the shaded band. To obtain the statistical errors in the jackknife analysis, we measure the minimum value of $B_4$ on each jackknife sample and use them for the error estimate, i.e. the minimum $\beta$ value fluctuates sample by sample. The value of $B_4$ at the CP in the $LT\to\infty$ limit, $b_4=1.604$, in Eq.~\eqref{eq:b4nu} is indicated by the black-dashed lines. 

Figure~\ref{fig:B4} shows that the results for different $LT$ cross around $\lambda=0.0007$, while the results at smaller $LT$ deviate from this trend; a clear deviation from the crossing point is visible even at $LT=8$ in the lower panel. 
Comparing this result to the one on the coarser lattice at $N_t=4$~\cite{Kiyohara:2021smr}, one finds that the deviation from the crossing at $N_t=6$ is more prominent at the same $LT$. This suggests that the violation of the FSS is more significant, and/or the effect of the mixing of energy-like observable in $\hat\Omega_{\rm R}$ described by Eq.~\eqref{eq:B4ansatz6} is stronger, on finer lattices.

\begin{table}[bt]
  \centering
  \caption{Results of the four-parameter fit based on Eq.~(\ref{eq:B4ansatz4}) for various combinations of $LT$. The first and second errors are the statistical and systematic ones, respectively, where the latter is estimated from the choices of $\lambda$ values. The last column shows $\chi^2$/dof of each fit.}
  \label{tab:fit}
  \begin{tabular}{lllll}
    \hline
    $LT$ & $b_4$ & $\lambda_{\rm c}\times10^4$ & $\nu$ & $\chi^2/{\rm dof}$ \\
    \hline
    $18,15,12$ & 1.6297(84)(6) & 7.048(52)(8) & 0.627(19)(5) & 0.40 \\
    $15,12,10$ &1.6294(81)(4) & 7.046(68)(4) & 0.626(17)(1) & 0.35 \\
    $18-10$ & 1.6295(55)(5) & 7.047(38)(3) & 0.621(11)(3) & 0.23 \\
    $18-9$ & 1.6331(41)(4) & 7.061(32)(7) & 0.616(87)(2) & 0.47 \\
    $18-8$ & 1.6418(32)(4) & 7.120(27)(8) & 0.616(87)(2) & 3.95 \\
    \hline
  \end{tabular}
\end{table}

\begin{table}[bt]
  \centering
  \caption{Results of the three-parameter fit based on Eq.~(\ref{eq:B4ansatz6}) for various combinations of $LT$.}
  \label{tab:fit2}
  \begin{tabular}{lll}
    \hline
    $LT$ & $\lambda_{\rm c}\times10^4$ & $\chi^2/{\rm dof}$ \\
    \hline
    $18,15,12$ & 6.986(32)(9) & 0.51 \\
    $15,12,10$ & 6.973(42)(5) & 0.83 \\
    $18-10$ & 6.984(25)(8) & 1.07 \\
    $18-9$ & 6.998(21)(5) & 4.23 \\
    $18-8$ & 7.035(19)(3) & 36.4 \\
    \hline
  \end{tabular}
\end{table}

\subsection{Fit analysis}
\label{sec:fit}

To determine the location of the CP and the critical exponent $\nu$, we perform the fit analyses of the lattice data based on Eqs.~\eqref{eq:B4ansatz4} and~\eqref{eq:B4ansatz6}.

First, we perform the four-parameter fit with Eq.~\eqref{eq:B4ansatz4}, where $b_4$, $c$, $\lambda_{\rm c}$, $\nu$ are the fitting parameters. For the data points to be fitted, we use $B_4$ at two values of $\lambda$, $\lambda_1$ and $\lambda_2$, for various combinations of $LT$, where the values of $\lambda_1$ and $\lambda_2$ are varied within the range $0.00067\le\lambda_1\le0.00069$ and $0.00071\le\lambda_2\le0.00073$ to estimate the systematic uncertainty associated to their choices. The values of $B_4$ at different $\lambda$'s are correlated in our analysis because their measurements are performed on the same gauge configurations through reweighting. We take the correlations into account by the correlated fits.

The results of the fits for $b_4,\lambda_{\rm c},\nu$ are summarized in Table~\ref{tab:fit} for various combinations of $LT$, where $(\lambda_1,\lambda_2)=(0.00069,0.00071)$ is employed for the central values. The first and second errors show the statistical and systematic errors, where the latter is estimated from the variations of $(\lambda_1,\lambda_2)$, which is well suppressed compared to the statistical error. The results for $LT=(18,15,12)$ and $LT=(15,12,10)$ are shown in the lower panel of Fig.~\ref{fig:B4} by the circle and square symbols with the double errors. 

The result in Table~\ref{tab:fit} shows that the value of the critical exponent $\nu$ is consistent with Eq.~\eqref{eq:b4nu}. However, $b_4$ has more than $2\sigma$ deviation from the $Z(2)$ value in Eq.~\eqref{eq:b4nu} even for the fit with $LT\ge12$. 

To investigate the possible mixing effects of energy-like observable to this result, we next perform the fits based on Eq.~(\ref{eq:B4ansatz6}). 
However, we found that the six-parameter fit, where $b_4$, $c$, $\lambda_{\rm c}$, $\nu$, $d$, $Y$ are the fitting parameters, is unstable with many local minima of $\chi^2$. We thus fixed $b_4$, $\nu$, $Y$ to the $Z(2)$ values in Eqs.~\eqref{eq:b4nu} and~\eqref{eq:Y} and performed the three parameter fit to determine $c$, $\lambda_{\rm c}$, and $d$. The results for $\lambda_{\rm c}$ for various combinations of $LT$ are shown in Table~\ref{tab:fit2}, where the meaning of errors is the same as Table~\ref{tab:fit} while the central value is set to $(\lambda_1,\lambda_2)=(0.00067,0.00071)$. The table shows that the fit works well with reasonable $\chi^2/{\rm dof}$ when $LT\ge 10$. The results for $LT=(18,15,12)$ and $LT=(15,12,10)$ are indicated in the lower panel of Fig.~\ref{fig:B4} by the diamond and triangle symbols at $B_4=b_4$ with horizontal error bars. The result for $LT=(18,15,12)$ is also shown in Fig.~\ref{fig:transition} by the circle symbol. 

\begin{figure*}
  \centering
  \includegraphics[width=0.32\textwidth]{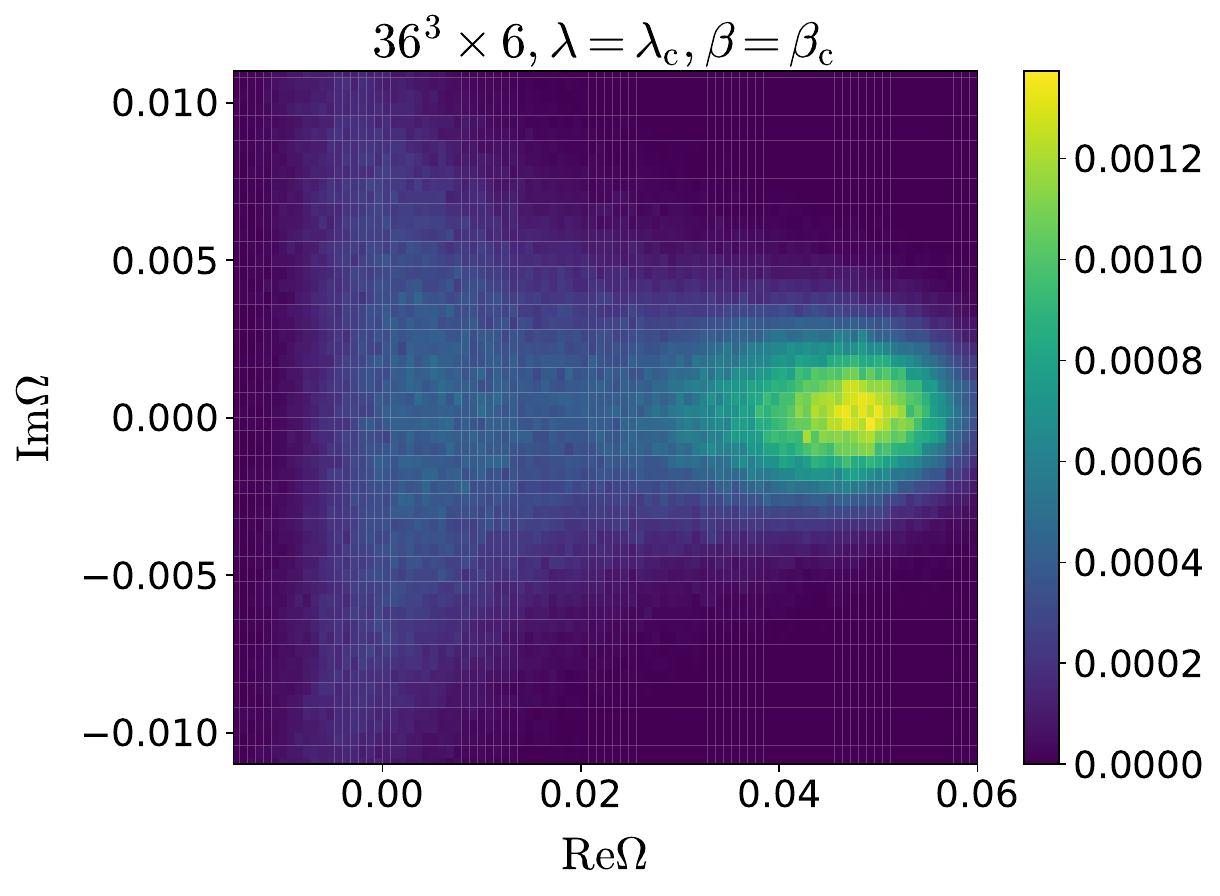}
    \includegraphics[width=0.32\textwidth]{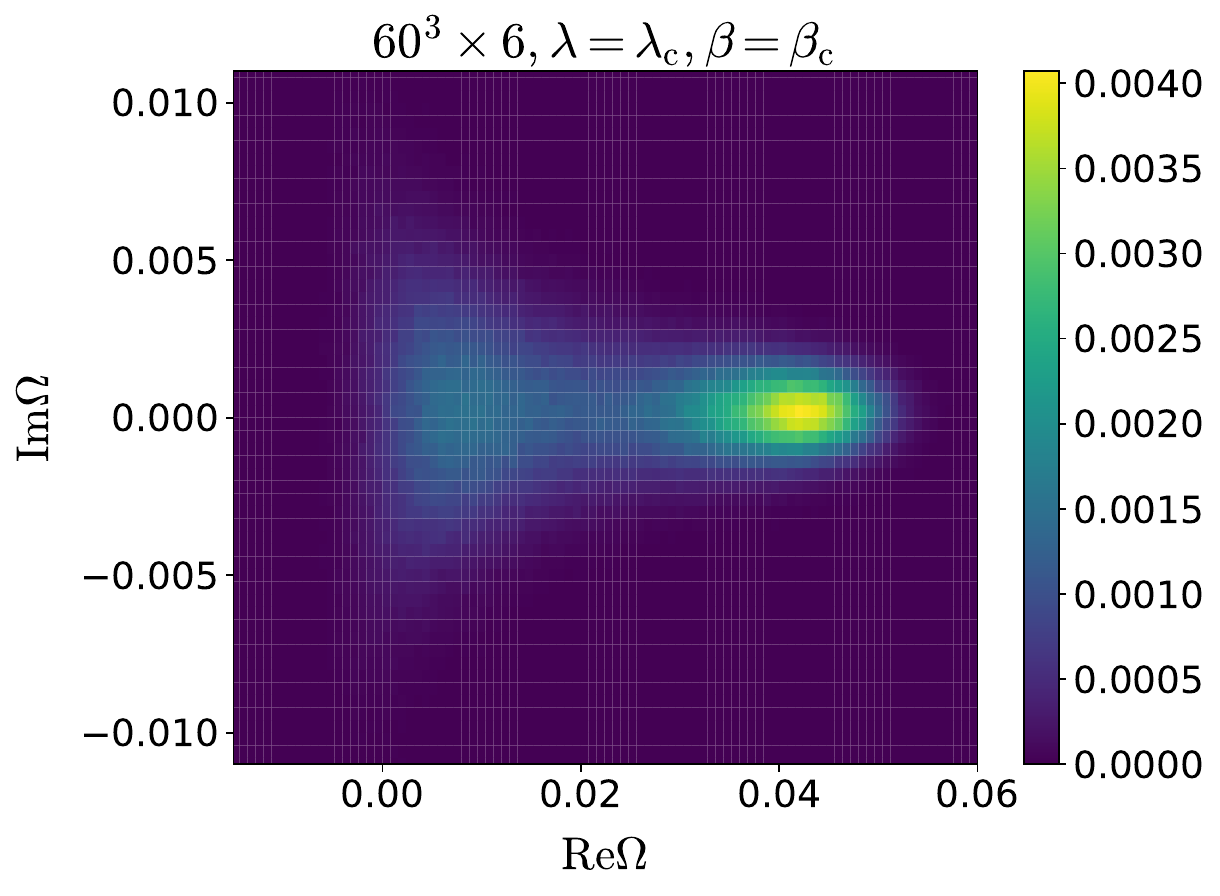}
    \includegraphics[width=0.32\textwidth]{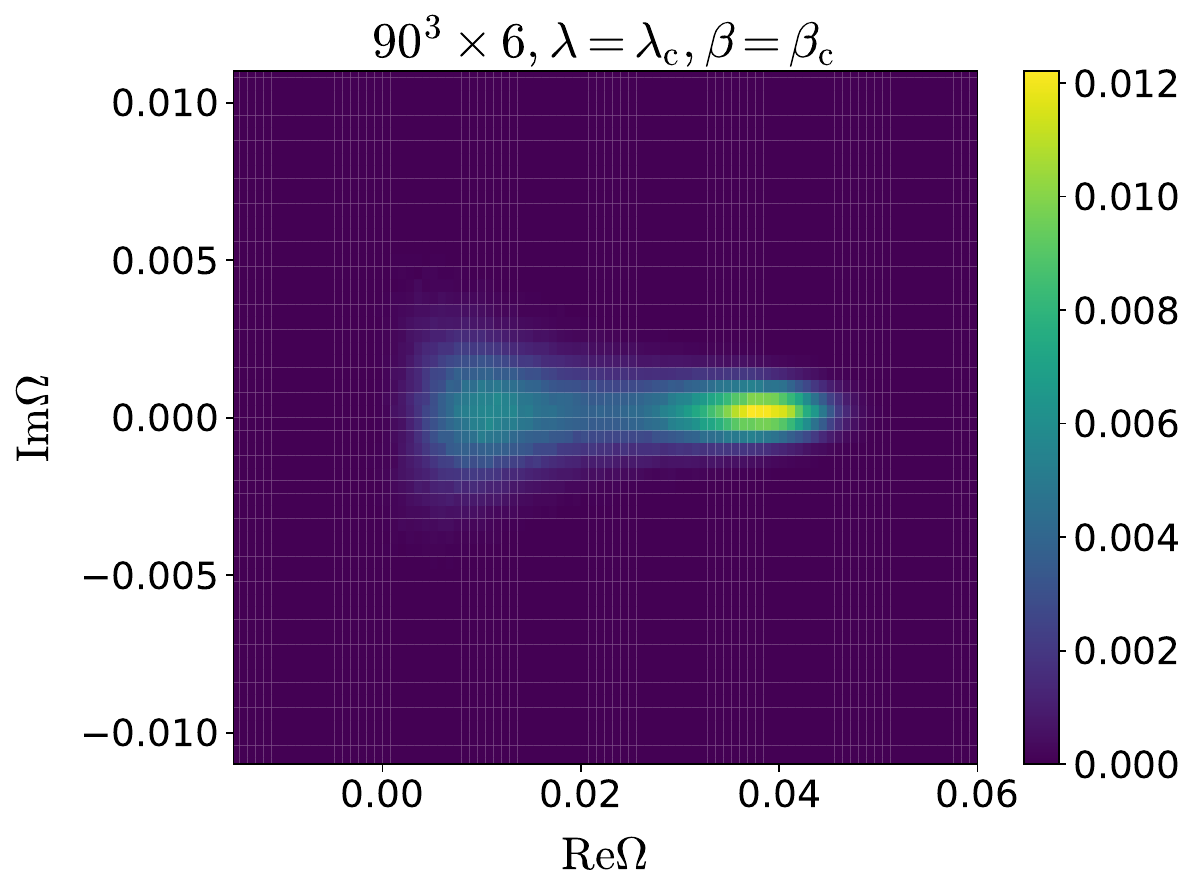}
  \caption{
    Two-dimensional histogram of $\hat\Omega$ on the complex plane at the critical point at $LT=6,10,15$.
  }
\label{fig:omegamap}
\end{figure*}

\subsection{Violation of the FSS}
\label{sec:omega}

Tables~\ref{tab:fit} and~\ref{tab:fit2} show that $\chi^2/{\rm dof}$ becomes unacceptably large when the numerical results for smaller $LT$ are included. This result indicates that the violation of the FSS shows up for $LT\le8$ in our data, presumably because the non-singular contribution to the free energy is not well suppressed there. 
Compared to Ref.~\cite{Kiyohara:2021smr}, the non-singular contribution seems to be amplified at the same $LT$ on the finer lattice.

To understand the origin of the violation of the FSS at small $LT$, we show in Fig.~\ref{fig:omegamap} the distribution of $\hat\Omega$ on the complex plane at $(\beta,\lambda)=(5.8817,0.0007)$, the parameter close to the CP, by the color-contour map for $LT=6,10,15$. From the left panel, one finds that the distribution has a triangular shape at $LT=6$, with distributions extending toward large $|{\rm Im}\hat\Omega|$ around ${\rm Re}\hat\Omega=0$. This behavior is clearly attributed to the remnant of the $Z(3)$ center symmetry at $\lambda=0$. As a result, the two peaks of the distribution are clearly asymmetric.

In the $Z(2)$ universality class, on the other hand, the scaling function has the rigorous $Z(2)$ symmetry; the magnetization is symmetric against the change of the sign in the Ising model. However, the breaking of this symmetry is manifest in the left panel of Fig.~\ref{fig:omegamap}, indicating the violation of the scaling behavior at $LT=6$. While the distribution approaches a symmetric one as $LT$ becomes larger, such violation is visible even on the right panel for $LT=15$.

The stronger violation of the FSS at $N_t=6$ than $N_t=4$ may be related to the strength of the first-order phase transition in SU(3) Yang-Mills (YM) theory corresponding to $\lambda=0$.
It is known that the latent heat in $SU(3)$ YM theory is large at $N_t=4$ owing to the lattice artifact~\cite{Shirogane:2016zbf,Shirogane:2020muc}. The large latent heat implies that a stronger external field is necessary to make the transition crossover. As a result, the CP is located at larger $\lambda$ at $N_t=4$, where the influence of the original $Z(3)$ symmetry at $\lambda=0$ is more suppressed. 

\begin{table}[bt]
    \caption{
    Parameters of the critical point at $N_t=6$ for $N_{\rm f}=1,2,3$. The central values are taken from the fit with Eq.~\eqref{eq:B4ansatz6}. The first errors are statistical, and the second are systematic ones due to the fit ansatz.}
    \label{tab:criticalNf}
    \centering
    \begin{tabular}{llll}
    \hline\hline
    $N_{\rm f}$ & $\lambda_{\rm c}$ & $\kappa_{\rm c}$ & $\beta_{\rm c}$ \\
   \hline
   1 & 0.0006100(27)($^{+46}_{-0}$) & 0.09624(7)($^{+12}_{-0}$) & 5.88355(5)($^{+0}_{-8}$)\\
   2 & 0.0006986(32)($^{+54}_{-0}$) & 0.08769(7)($^{+11}_{-0}$) & 5.88180(5)($^{+0}_{- 9}$) \\
   3 & 0.0007491(36)($^{+59}_{-0}$) & 0.08292(7)($^{+11}_{-0}$) & 5.88055(6)($^{+0}_{-10}$) \\
   \hline\hline
    \end{tabular}
\end{table}

\subsection{Location of CP: $N_{\rm f}$ dependence}

So far, we have discussed the numerical results for the case of $N_{\rm f}=2$. The generalization of these results to other $N_{\rm f}$ and non-degenerate cases is straightforward as the $N_{\rm f}$-dependence of the HPE is explicitly known.
In Table~\ref{tab:criticalNf}, we summarize the location of the CP for $N_{\rm f}=1,2,3$.
The fitting result using Eq.~\eqref{eq:B4ansatz6} with $LT=(18,15,12)$ is employed as the central value, and the difference from the fit with Eq.~\eqref{eq:B4ansatz4} is shown by the second error as an estimate of the systematic error due to the fit ansatz.

The location of the CP in the heavy-quark QCD with $N_{\rm f}=2$ at $N_t=6$ has been investigated in Ref.~\cite{Cuteri:2020yke} on smaller lattices with the same gauge and quark actions as ours. In this study, the value of $\kappa$ at the CP is estimated as $\kappa_{\rm c}=0.0877(9)$. This value agrees with the one in Table~\ref{tab:criticalNf}, while the error is significantly suppressed in our analysis.

\section{Summary}
\label{sec:summary}

In this study, we have performed the Binder-cumulant analysis of the CP in the heavy-quark QCD. We extended our previous analysis at $N_t=4$ to finer lattices with $N_t=6$. To see the proper scaling behavior, numerical analyses have been performed on large lattices up to the aspect ratio $LT=18$. The gauge configurations are generated at the LO action in the HPE, and the effects of higher-order terms are incorporated by the reweighting method. Contributions of the NLO are treated exactly, and yet higher-order terms are incorporated effectively up to virtually infinite order. 
We have succeeded in realizing high-precision analysis of the Binder cumulant on large spatial volumes while suppressing the truncation errors of the HPE. Precision results for the location of the CP at $N_t=6$ have been obtained from these results for $N_{\rm f}=1,2,3$.
We also found that the violation of the scaling behavior due to finite volume is more prominent at $N_t=6$ than our previous study at $N_t=4$. From the distribution of the Polyakov loop in the complex plane, we argued that the remnant of the $Z(3)$ symmetry at $\lambda=0$ affects this result.

\section*{acknowledgment}

We thank Yasumichi Aoki, Hiroshi Suzuki, Takashi Umeda, and Naoki Wakabayashi for useful discussions. 
This work was supported in part by JSPS KAKENHI (Nos.~JP19H05598, JP20H01903, JP21K03550, JP22K03593, JP22K03619, JP23H04507, JP24K07049), and by the Center for Gravitational Physics and Quantum Information (CGPQI) at Yukawa Institute for Theoretical Physics.
Numerical simulations were done in part under the HPCI System Research projects (Project ID: hp200013, hp200089, hp210012, hp210039, hp220020, hp220024), the Research proposal-based use at the Cybermedia Center, Osaka University, and the Multidisciplinary Cooperative Research Program of the Center for Computational Sciences, University of Tsukuba.

\appendix

\section{Convergence of HPE}
\label{sec:convergence}

In this appendix, we discuss the convergence of the HPE. 

\begin{figure}
  \centering
    \includegraphics[width=0.4\textwidth]{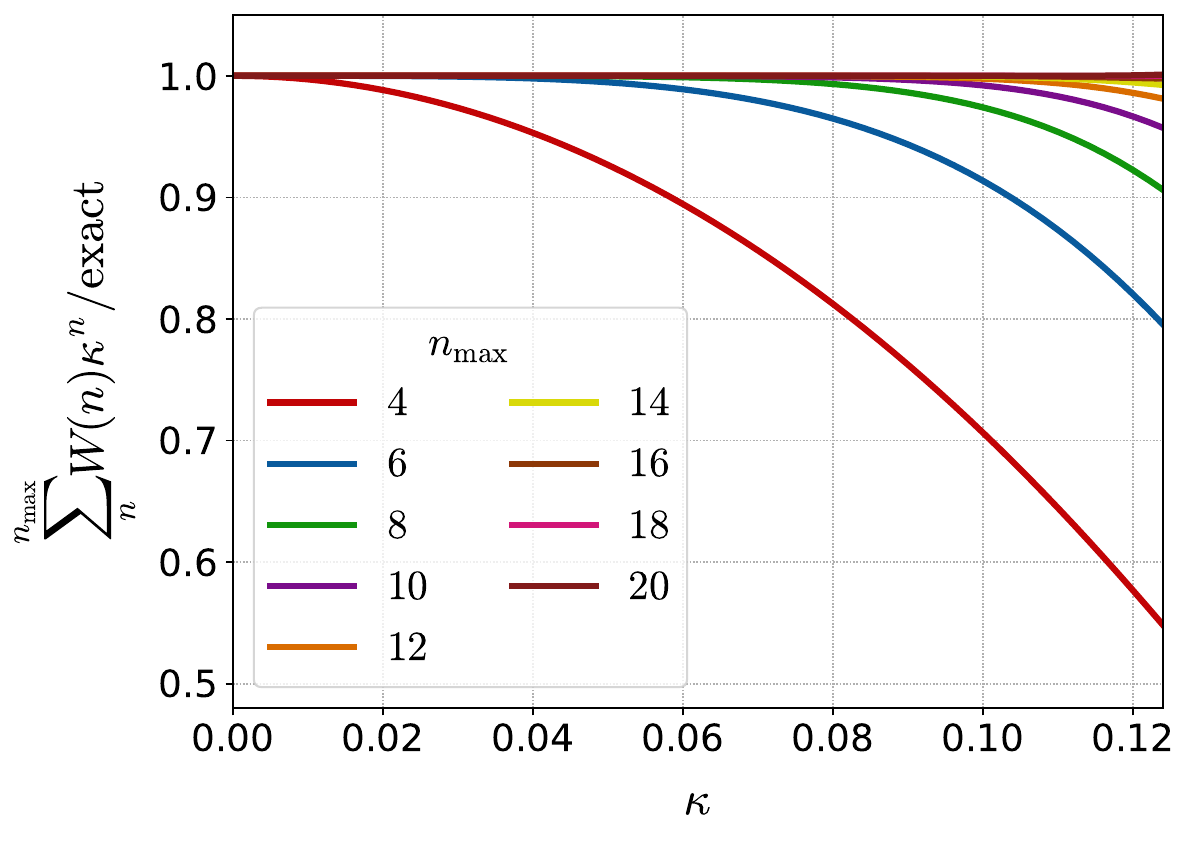}
    \includegraphics[width=0.4\textwidth]{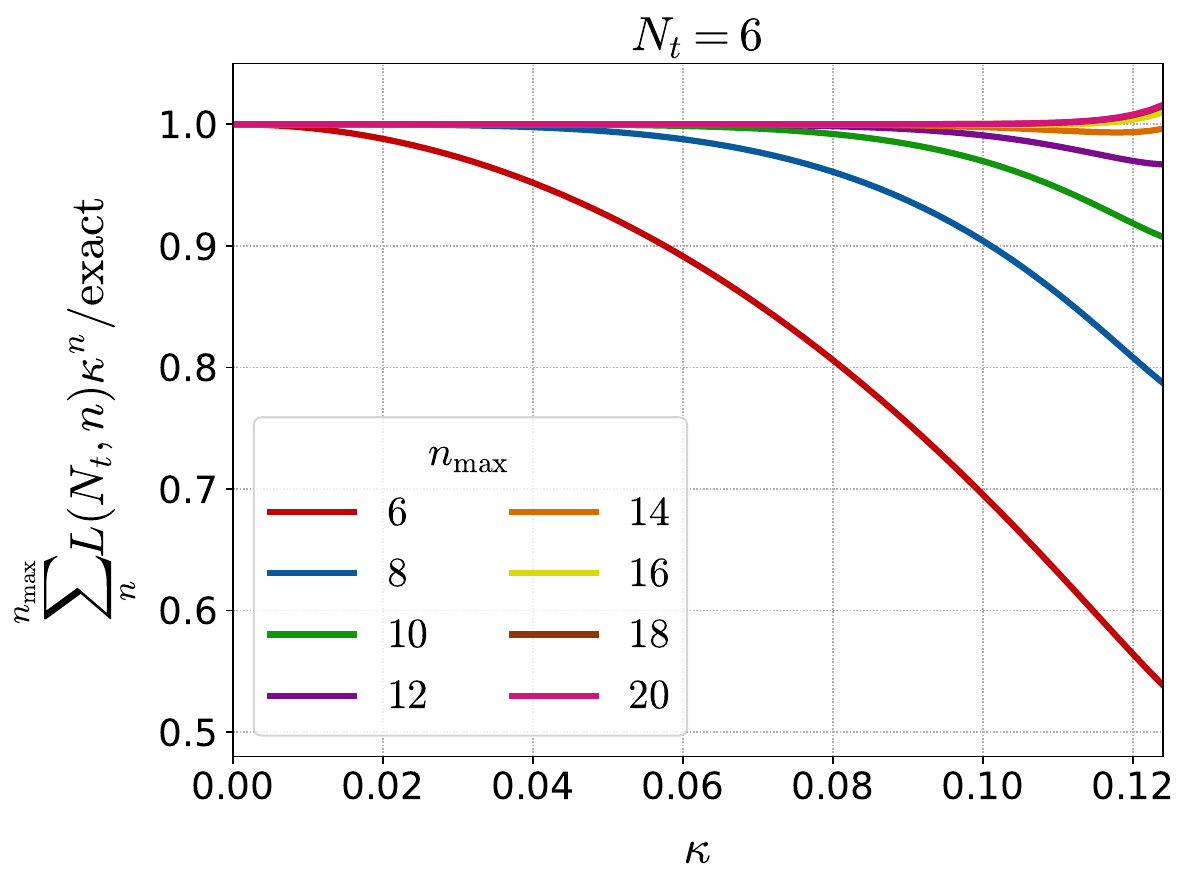}
  \caption{
    Convergence of the HPE in the weak-coupling limit. The upper (lower) panel shows the contributions of the Wilson (PLT) loops up to $n$th order to the action $S$ normalized by the exact value.
  }
\label{fig:truncation}
\end{figure}

\subsection{HPE for free-fermion system}
\label{sec:free}

In this subsection, we study the magnitudes of higher order terms of the HPE for the free lattice fermion, which is obtained by substituting $U_{x,\mu}=1$ into Eq.~(\ref{eq:Sq}). As discussed in Ref.~\cite{Wakabayashi:2021eye}, in this case one can calculate the values of $\hat W(n)$ and $\hat L(N_t,n)$ analytically, which are denoted by $W_0(n)$ and $L_0(N_t,n)$ in the text.

For the free Wilson fermions, the hopping term~(\ref{eq:B}) is given by~\cite{Wakabayashi:2021eye}
\begin{align}
    b_{xy}=\sum_{\mu=1}^4 \left[ (1-\gamma_{\mu})\,\delta_{y,x+\hat{\mu}} + (1+\gamma_{\mu})\,\delta_{y,x-\hat{\mu}} \right] ,
  \label{eq:b}
\end{align}
where the Dirac-spinor and color indices are suppressed for notational simplicity.
To calculate $W_0(n)$, one may consider infinitely large lattice and Fourier transform the spatial coordinates of Eq.~(\ref{eq:b}). This procedure leads to 
\begin{align}
    W_0(n) 
    =& -\frac1n {\rm Tr}[b^n]
    \notag \\
    =&
    - \frac{N_{\rm c}}n \int_0^{2\pi} \frac{d^4\varphi}{(2\pi)^4} {\rm tr} [\tilde{b}(\varphi)^n] ,
    \label{eq:W=int}
\end{align}
with 
\begin{align}
    \tilde b(\varphi) 
    =&\ \tilde b(\varphi_1,\varphi_2,\varphi_3,\varphi_4) 
    \notag \\
    =&\ \sum_{\mu=1}^4 \left[ (1-\gamma_{\mu})\,e^{i\varphi_\mu} + (1+\gamma_{\mu})\,e^{-i\varphi_\mu} \right] .
    \label{eq:b(phi)}
\end{align}
Notice that the PLT loops do not contribute to Eq.~\eqref{eq:b(phi)} since both $N_s$ and $N_t$ are taken to be infinite.

To calculate $L_0(N_t,n)$ we perform the same procedure on the lattice of size $N_s^3\times N_t$, which leads to 
\begin{align}
    & W_0(n) + L_0(N_t,n)
    \notag \\
    &= -\frac{N_c}{N_{\rm site}n} \sum_{k_1,k_2,k_3=1}^{N_s} \sum_{k_4=1}^{N_t} {\rm tr} \Big[ \tilde{b}\big(\frac{2\pi k_1}{N_s},\frac{2\pi k_2}{N_s},\frac{2\pi k_3}{N_s},\frac{2\pi k_4}{N_t}\big)\Big].
    \label{eq:W+L=sum}
\end{align}
Then, substracting Eq.~\eqref{eq:W=int} from Eq.~\eqref{eq:W+L=sum} gives $L_0(N_t,n)$.
Note that Eq.~(\ref{eq:W+L=sum}) is equivalent to Eq.~(\ref{eq:W=int}) for $n<N_s$ and $n<N_t$. 

For the free-quark system, the quark determinant is given by
\begin{align}
    -{\rm ln}{\rm det}M 
    = \sum_n^\infty W_0(n)\kappa^n + \sum_n^\infty L_0(N_t,n)\kappa^n.
    \label{eq:lndetM0}
\end{align}
To see the truncation error of the HPE in Eq.~\eqref{eq:lndetM0}, in Fig.~\ref{fig:truncation} we show 
\begin{align}
    \sum_n^{n_{\rm max}} W_0(n)\kappa^n, \qquad \sum_n^{n_{\rm max}} L_0(N_t,n) ,
    \label{eq:sumWL}
\end{align}
normalized by the exact value for various $n_{\max}$.
In the lower panel, we show the case of $N_t=6$. From the figure, one sees that Eq.~\eqref{eq:sumWL} approaches the exact value as $n_{\rm max}$ becomes larger. As discussed in Sec.~\ref{sec:result}, the value of $\kappa$ at the critical point is $\kappa_{\rm c}=0.0877$. At this $\kappa$, the HPE at NLO, i.e. $n_{\rm max}=6$ and $8$ for the Wilson and PLT loops, respectively, is already beyond $95\%$ of the exact one. This result suggests that the analysis at NLO gives a good approximation around the CP at $N_t=6$. The inclusion of one more higher-order contribution gives an almost perfect approximation. 

As discussed in Ref.~\cite{Wakabayashi:2021eye}, the convergence of the HPE in the interacting systems is faster than the free-fermion case, since non-unit link variables $U_{x,\mu}$ tends to suppress the higher order terms more significantly.

\begin{figure}
  \centering
    \includegraphics[width=0.4\textwidth]{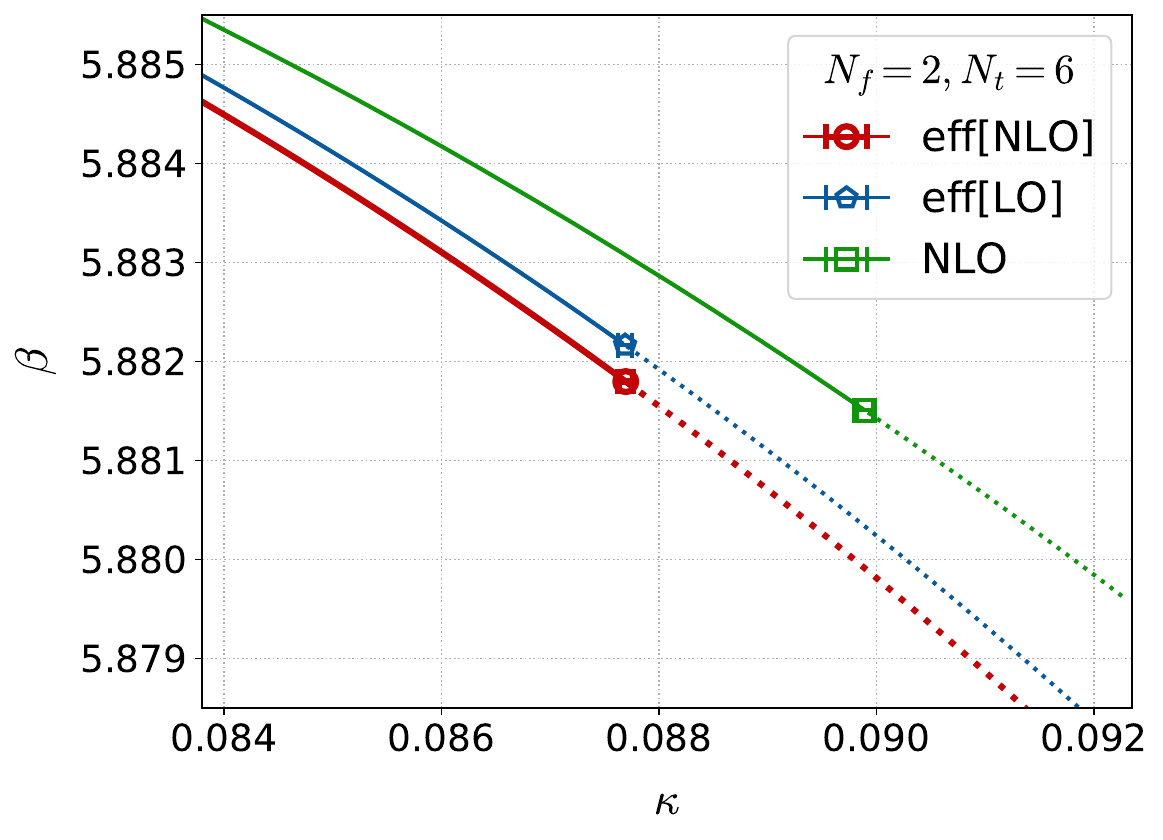}
  \caption{
  Transition line and the CP obtained for the action Eq.~\eqref{eq:SnDnL3} (eff[NLO]), Eq.~\eqref{eq:SnDnL2} (eff[LO]), and the action at NLO. The solid and dashed lines show the first-order transition and crossover.
  The Eff-NLO result is the same as that shown in Fig.~\ref{fig:transition}.
  }
\label{fig:phase_comp}
\end{figure}

\subsection{Location of the CP in heavy-quark QCD}
\label{sec:comparison}

In Sec.~\ref{sec:effective}, we introduced two methods to effectively incorporate higher-order terms of the HPE with the actions~\eqref{eq:SnDnL2} and~\eqref{eq:SnDnL3}. 
In this subsection, we compare the numerical results obtained by these methods, as well as the analysis at NLO.

In Fig.~\ref{fig:phase_comp}, we show the transition line as a function of $\kappa$ and the CP determined adopting the actions Eq.~\eqref{eq:SnDnL3} (eff[NLO]) and Eq.~\eqref{eq:SnDnL2} (eff[LO]), together with the results at NLO, where the transition lines are obtained from the minimum of $B_4$ at $LT=15$. The statistical errors are smaller than the thickness of the lines. The CPs are determined by the same analysis as for Table~\ref{tab:criticalNf}.

From the figure, we find that, though the NLO calculation should contain about $95\%$ contributions to the action as discussed in the previous subsection, the incorporation of higher-order terms causes a shift of the transition line and the CP towards smaller $\kappa$ compared to the NLO results.
Thus, the incorporation of higher-order effects is important for their precise determination.
On the other hand, the two alternatives to incorporate higher-order terms, eff[NLO] and eff[LO], give similar results. 
As discussed in Sec.~\ref{sec:DL}, these results are stable under variations of $n_{\rm W}$ and $n_{\rm L}$. These suggest that the truncation error is well suppressed in these methods. As the eff[NLO] should give the most reliable results among them, we use it in the analyses in the main text.

\bibliographystyle{apsrev4-1}
\bibliography{refs.bib}

\end{document}